\documentclass[preprint,prc,aps,showpacs]{revtex4-1}

\usepackage{graphicx}
\usepackage{xcolor}

\usepackage{amsmath} %pacote para usar o comando \begin{align} e etc...
\usepackage{amssymb} %pacote para usar o comando \begin{align} e etc...
\usepackage{subfig}   %Permite utilizar subfiguras 
\usepackage{multirow}
\def\subfigure{\subfloat}  

\newcommand{\bra}[1]{\langle #1|}
\newcommand{\ket}[1]{|#1\rangle}
\newcommand{\braket}[2]{\langle #1|#2\rangle}
\usepackage[ansinew]{inputenc}
\begin{document}
%\selectlanguage{english}
\title{A QCD sum rules calculation of the $\eta_c D^* D$ and $\eta_c D_s^* D_s$  form factors and strong coupling constants}

\author{B. Os\'orio Rodrigues $^a$}\email{brunooz@uerj.br}
\author{M. E. Bracco $^b$}
\author{C. M. Zanetti $^b$ } 

\affiliation{$^a$ Instituto de Aplica\c{c}\~ao Fernando Rodrigues da Silveira, Universidade do Estado do Rio de 
Janeiro, Rua Santa Alexandrina 288, 20261-232, Rio de Janeiro, RJ, Brazil. }
\affiliation{$^b$ Faculdade de Tecnologia, Universidade do Estado do Rio de Janeiro, 
Rod. Presidente Dutra Km 298, P\'olo Industrial, 27537-000, Resende, RJ, Brazil.}

\begin{abstract}
We use the QCD sum rules  for the three point correlation functions to
compute the  strong coupling constants  of the meson  vertices $\eta_c
D^*  D$  and  $\eta_c  D_s^*   D_s$.   We  consider  perturbative  and
non-perturbative contributions,  working up  to dimension five  on the
OPE.  The vertices were studied considering that each one of its three
mesons  are off-shell  alternately.  The  vertex coupling  constant is
evaluated  through  the  extrapolation  of the  three  different  form
factors.   The  results  obtained   for  the  coupling  constants  are
$g_{\eta_c  D^*  D}  =   5.23^{+1.80}_{-1.38}$  and  $g_{\eta_c  D_s^*
  D_s}=5.55^{+1.29}_{-1.55}$.

\end{abstract}

\maketitle

\section{Introduction}

In  this paper,  we  add two  more calculations  of  form factors  and
coupling constants to the set of charmonium processes: the $\eta_c D^*
D$ and  $\eta_c D_s^* D_s$ vertices.   We use here the  same technique
described  in  the review  of  charmonium  form factors  and  coupling
constants, that were developed  by our group~\cite{Bracco:2011pg}.  Our
technique,    which     uses    the     QCD    Sum     Rules    method
(QCDSR)~\cite{1979NuPhB.147..385S,Shifman:1978by}, was  polished by us
and  allowed  to  improve  the  results with  better  control  of  the
uncertainties that are typical of the QCDSR method.

The  technique  of  our  group  takes advantage  of  the  form  factor
calculation for  all the different  off-shell cases of a  same vertex.
The  form factor,  which  is  a function  of  the squared  transferred
momentum ($q^2$), can be very different  when this momentum is the one
that  represents  the light  meson  or  if  it  is the  momentum  that
represents the heavy meson of  the process. Nevertheless, the coupling
constant of the  process is unique for the vertex,  independent of the
meson  that is  considered  off-shell.  Our  objective doing  multiple
calculations  of  the form  factors  to obtain an  unique  coupling
constant  is to  minimize the  uncertainties of  the QCDSR  technique,
demanding that all form factors of  a same vertex converge in the same
coupling  constant.  In  this work,  we make  the calculations  of the
three  possible form  factors  of  the three  mesons  of the  vertices
$\eta_c  D^* D$  and $\eta_c  D_s^* D_s$  obtaining a  unique coupling
constant for each vertex.

The  $\eta_c D^*  D$ and  $\eta_c  D_s^* D_s$  coupling constants  are
necessary  in some  calculations  of decay  processes, with  different
motivations.  In  the paper  of  Qian  Wang {\it  et.
    al.}~\cite{Wang2012364}, the properties of the $\eta_c \rightarrow
  V V$  decay are studied  as an alternative  process for the  test of
  intermediate meson  loop transition.  The form  factors and coupling
  constant are included in the  calculation. The results are sensitive
  to the  form factor  parameters and the  coupling constant  used are
  obtained in the quiral and heavy quark limit.

In     studies    of     $e^+e^-     \rightarrow    J/\psi     \eta_c$
processes~\cite{Zhang:2008ab},   where   it   was   investigated   the
intermediate mesons  loop contribution with $D$  mesons,  monopolar
form factors are used and for the coupling constant it was adopted the
relation in  the heavy  quark limit.

In general, model predictions for the form factors can vary as much
as 30\% which in turn imply large uncertainties in the branching
ratios. Besides that,  more than two parameters can be necessary
  in the form factors without a clear reason. 

We  will  take  advantage  of  the  similarities  between  the  mesons
$D^{(*)}$ and $D_s^{(*)}$ in order to also calculate in this paper the
$g_{\eta_c D_s^* D_s}$ coupling constant for the $\eta_c$, $D_s^*$ and
$D_s$  off-shell  cases  and  we  will  compare  both  final  coupling
constants  for each  vertex with  each  other according  to the  SU(4)
symmetry.

\section{The three point correlation function}

In  the  QCDSR  approach,  the coupling  constants  for  three  mesons
vertices can be  evaluated through the computation of  the three point
correlation function~\cite{Bracco:2011pg}.   The correlation functions
contain information about  the quantum numbers of the  mesons that are
part of the vertex, and can be computed in two distinct manners: using
hadronic degrees  of freedom on  the phenomenological side,  and using
quarks and gluons  degrees of freedom on the OPE  side.  The QCDSR are
obtained by applying the  quark-hadron duality principle, which allows
the matching of both sides of  the correlation function, by applying a
double Borel transformation, thus obtaining an analytic expression for
the vertex  form factor.  In order  to compute the numerical  value of
the coupling  constant, the form  factor is extrapolated to  the meson
pole $Q^2=-m^2$, where $m$ is the meson mass.  In the case of vertices
with  three distinct  mesons, this  procedure allows  to obtain  three
distinct  vertex coupling  constants,  one for  each off-shell  meson.
However, the  vertex is the  same regardless of which  off-shell meson
being considered,  hence the three constant  couplings obtained should
be   equal.    Using   this   approach,  that   were   introduced   on
Ref.~\cite{Bracco:2011pg},  the  uncertainties  are minimized  and  we
obtain only one coupling constant for the vertex.

Considering the case of the vertex $\eta_c \mathcal{D}^* \mathcal{D}$,
where $\mathcal{D}^{(*)}  = \left  (D^{(*)},D_s^{(*)}\right )$,  it is
possible  to   set  up  the  following   three  different  correlation
functions, $\Gamma^{(M)}_\mu(p,p')$, where $M$  is the off-shell meson
($M = \eta_c,\,\mathcal{D}^*,\,\mathcal{D}$):

\begin{eqnarray}
\Gamma^{(\eta_c)}_\mu(p,p')      &=&       \int      \bra{0'}      T\{
j^{\mathcal{D}}_5(x)j^{\eta_c\dagger}_5(y)j^{\mathcal{D}^*\dagger}_\mu(0)
\}\ket{0'}    e^{ip'x}e^{-iqy}d^4x     d^4y\,,    \label{eq:pietacoff}
\\   \Gamma^{(\mathcal{D}^*)}_\mu(p,p')   &=&    \int   \bra{0'}   T\{
j^{\eta_c}_5(x)j^{\mathcal{D}^*\dagger}_\mu(y)j^{\mathcal{D}}_5(0)
\}\ket{0'}                                        e^{ip'x}e^{-iqy}d^4x
d^4y\,,  \label{eq:piDEstoff}\\  \Gamma^{(\mathcal{D})}_\mu(p,p')  &=&
\int                            \bra{0'}                           T\{
j^{\eta_c}_5(x)j^{\mathcal{D}\dagger}_5(y)j^{\mathcal{D}^*}_\mu(0)
\}\ket{0'} e^{ip'x}e^{-iqy}d^4x d^4y\,, \label{eq:piDoff}
\end{eqnarray}
where $\ket{0'}$ is  the non-trivial QCD vacuum,  $q = p' -  p$ is the
transferred four-momentum,  $j^{\eta_c}_5$, $j^{\mathcal{D}^*}_\mu$ and
$j^{\mathcal{D}}_5$ are respectively the $\eta_c$, $\mathcal{D}^*$ and
$\mathcal{D}$ mesons' interpolating currents.

The   first   step   to   apply   the  QCDSR   is   to   compute   the
phenomenological and OPE sides of the
Eqs.~(\ref{eq:pietacoff})-(\ref{eq:piDoff}).

\subsection{The phenomenological side}
We initiate  this section considering  the Lagrangian of  the hadronic
process that is  required to compute the phenomenological  side of the
QCDSR. For the vertex $  \eta_c \mathcal{D}^* \mathcal{D}$, we use the
following         expression         for        the         Lagrangian
($\mathcal{L}$)~\cite{Wang2012364}:
\begin{align}
\mathcal{L}_{\eta_c   \mathcal{D}^*   \mathcal{D}}  =   -i   g_{\eta_c
  \mathcal{D}^*     \mathcal{D}}\left     [     \mathcal{D}^{*+\alpha}
  (\partial_\alpha   \mathcal{D}^-\eta_c   -  \partial_\alpha   \eta_c
  \mathcal{D}^-)  +   \mathcal{D}^{*-\alpha}  (\partial_\alpha  \eta_c
  \mathcal{D}^+ - \eta_c \partial_\alpha \mathcal{D}^+) \right ]\,.
\label{eq:lagrangiana}
\end{align}

From the  former expression, the  following vertices are  obtained for
the  cases  of  the  off-shell mesons  $\eta_c$,  $\mathcal{D}^*$  and
$\mathcal{D}$, respectively:
\begin{eqnarray}
\braket{\mathcal{D}^*(p)\eta_c(q)}{\mathcal{D}(p')}        &=&       i
g^{(\eta_c)}_{\eta_c                                     \mathcal{D}^*
  \mathcal{D}}(q^2)\epsilon^\alpha(p)(2p'_\alpha - p_\alpha) \,,
\end{eqnarray}
\begin{eqnarray}
\braket{\mathcal{D}(p)\mathcal{D}^*(q)}{\eta_c(p')}        &=&       i
g^{(\mathcal{D}^*)}_{\eta_c                              \mathcal{D}^*
  \mathcal{D}}(q^2)\epsilon^\alpha(q)(p'_\alpha + p_\alpha) \,,
\end{eqnarray}
\begin{eqnarray}
\braket{\mathcal{D}^*(p)\mathcal{D}(q)}{\eta_c(p')}        &=&       i
g^{(\mathcal{D})}_{\eta_c                                \mathcal{D}^*
  \mathcal{D}}(q^2)\epsilon^{\alpha}(p)(2p'_\alpha - p_\alpha) \,,
\end{eqnarray}
where $\epsilon^{\alpha}$  is the  polarization vector  and $g_{\eta_c
  \mathcal{D}^* \mathcal{D}}^{(M)}(q^2)$ are the vertices' form factors
with the off-shell mesons $M = \eta_c,\,\mathcal{D}^*,\,\mathcal{D}$.

In order to obtain an expression with hadronic degrees of freedom
for the phenomenological side, the  intermediate states of the mesons
are       inserted      in       the      correlation       functions,
Eqs.~(\ref{eq:pietacoff})-(\ref{eq:piDoff}),
and in which the following matrix elements are used:
\begin{eqnarray}
\bra{P(q)}j^P_5\ket{0}   &=&   \bra{0}    j^P_5   \ket{P(q)}   =   f_P
\frac{m_P^2}{m_{q_1}  + m_{q_2}}\\  \bra{V(q)}  j_\mu^{V} \ket{0}  &=&
f_{V} m_{V} \epsilon^*_\mu(q)\,.
\end{eqnarray}
where $P$ is a pseudo-scalar meson ($P = \eta_c, \mathcal{D}$), $V$ is
a vector-meson ($V = \mathcal{D}^*$),  $q$ is the four-momentum of the
respective meson, $m_{P,V}$ is the  meson mass, $f_{P,V}$ is the meson
decay  constant, and  $m_{q_1}$  and $m_{q_2}$  are quark  constituent
masses of the  meson $P$.

The expressions  thus obtained  for the  correlation functions  on the
phenomenological side are:
\begin{eqnarray}
\Gamma^{phen   (\eta_c)}_\mu   &   =   &   C\frac{g^{(\eta_c)}_{\eta_c
    \mathcal{D}^*   \mathcal{D}}(q^2)\left   [(m^2_{\mathcal{D}^*}   +
    m^2_\mathcal{D} -  q^2 )p_\mu -  2m^2_{\mathcal{D}^*}p'_\mu \right
]}{(p^2   -  m_{\mathcal{D}^*}^2)(p'^2   -  m_{\mathcal{D}}^2)(q^2   -
  m_{\eta_c}^2)  } +  h. r.  \,,\label{eq:fenomEtacoff}\\ \Gamma^{phen
  (\mathcal{D}^*)}_\mu   &   =  &   C\frac{g^{(\mathcal{D}^*)}_{\eta_c
    \mathcal{D}^*    \mathcal{D}}(q^2)\left   [(m^2_{\mathcal{D}}    -
    m^2_{\mathcal{D}^*}   -  m^2_{\eta_c})p_\mu   +  (m^2_{\eta_c}   -
    m^2_{\mathcal{D}}  - m^2_{\mathcal{D}^*})p'_\mu  \right ]}{(p^2  -
  m_{\mathcal{D}}^2)(q^2 - m_{\mathcal{D}^*}^2)(p'^2 - m_{\eta_c}^2) }
+ h. r. \,,\label{eq:fenomDEstoff}\\ \Gamma^{phen (\mathcal{D})}_\mu &
=        &       C\frac{g^{(\mathcal{D})}_{\eta_c        \mathcal{D}^*
    \mathcal{D}}(q^2)\left  [(m^2_{\mathcal{D}^*}   +  m^2_{\eta_c}  -
    q^2)p_\mu   -   2m^2_{\mathcal{D}^*}p'_\mu    \right   ]}{(q^2   -
  m_{\mathcal{D}}^2)(p^2 - m_{\mathcal{D}^*}^2)(p'^2 - m_{\eta_c}^2) }
+ h. r. \,,\label{eq:fenomDoff}
\end{eqnarray}
where  $h.r.$  are  the  contributions from  the  resonances  and  the
continuum and $C$ is defined as:
\begin{align}
C   =  \frac{f_{\eta_c}f_{\mathcal{D}^*}f_{\mathcal{D}}   m^2_{\eta_c}
  m^2_{\mathcal{D}}}{2m_{\mathcal{D}^*} m_c (m_c + m_q)},
\label{eq:C}
\end{align}
with $m_q =  (m_u, m_d, m_s)$, depending  if $\mathcal{D}$ corresponds  
to the meson $D$ or $D_s$.

\subsection{The OPE side}
The OPE side is calculated  by inserting the interpolating currents in
terms         of          quark         fields          in         the
Eqs.~(\ref{eq:pietacoff})-(\ref{eq:piDoff}). In this  work, we use the
following currents:
$$j_5^{\eta_c}   =   i
\bar{c}\gamma_5 c,$$  $$j_\mu^{\mathcal{D}^*+} = \bar{q}\gamma_\mu  c,$$
$$j_5^{\mathcal{D}+} =  i \bar{q}\gamma_5  c,$$
where  $q$ is  a light
quark ($q  = u,d,s$) whose flavor  corresponds to the light  quark of
the given open charm meson, $\mathcal{D}=(D^{(*)\pm},D_s^{(*)\pm})$.

The  OPE  side  is  regarded  as an  ordinate  series  by  the  Wilson
operators,   obtained   from   the  expansion   of   the   correlation
function.   The  series   is  dominated   by  the   perturbative  term
($\Gamma^{\text{pert}(M)}_\mu$),  followed   by  the  non-perturbative
contributions to the correlator ($\Gamma^{\text{non-pert}(M)}_\mu$):
\begin{align}
\Gamma^{\text{OPE}(M)}_\mu = \Gamma^{\text{pert}(M)}_\mu + \Gamma^{\text{non-pert}(M)}_\mu \,.
\label{eq:piladodaqcdgeral}
\end{align}
In the  calculation of  form factors for  three mesons'  vertices, such
expansion usually presents a fast convergence of the series and can be
truncated   after  a   few   terms.   In   this   work,  we   consider
non-perturbative contributions  on the  OPE up  to fifth  order, which
includes the quark-gluon mixed condensates:
\begin{align}
\Gamma^{\text{non-pert}}_\mu  = \Gamma^{\langle\bar{q}q\rangle}_\mu  +
\Gamma^{m_q\langle\bar{q}q\rangle}_\mu    +     \Gamma^{\langle    g^2
  G^2\rangle}_\mu + \Gamma^{\langle \bar{q}g\sigma  G q \rangle}_\mu +
\Gamma^{m_q\langle \bar{q}g\sigma G q \rangle}_\mu\,.
\label{eq:nonpertcontrib}
\end{align}

\begin{figure}[ht!]
\centering
\subfigure[]{\includegraphics[width=0.29\linewidth]{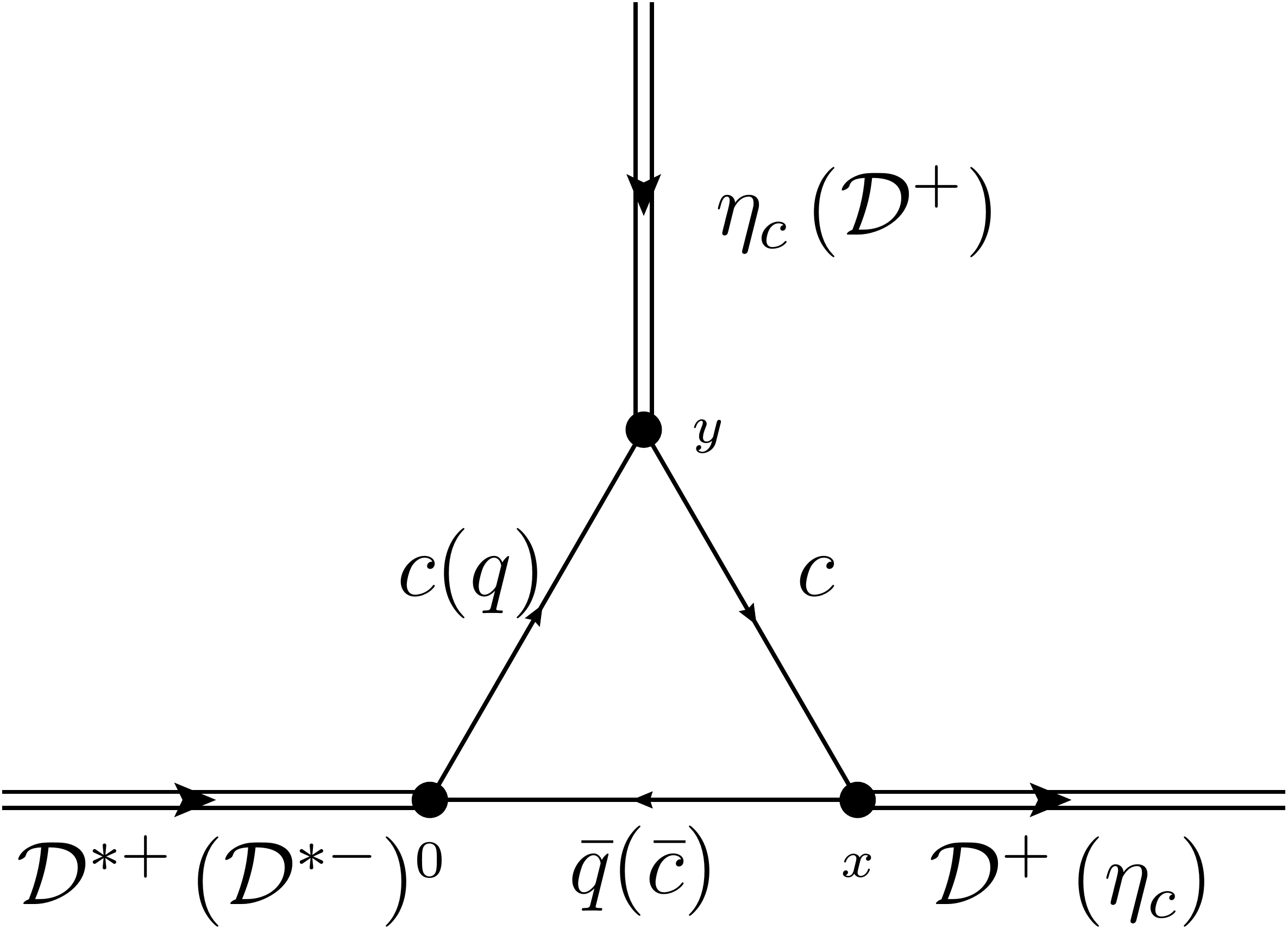} \label{subfig:a}}\quad
 \subfigure[]{\includegraphics[width=0.29\linewidth]{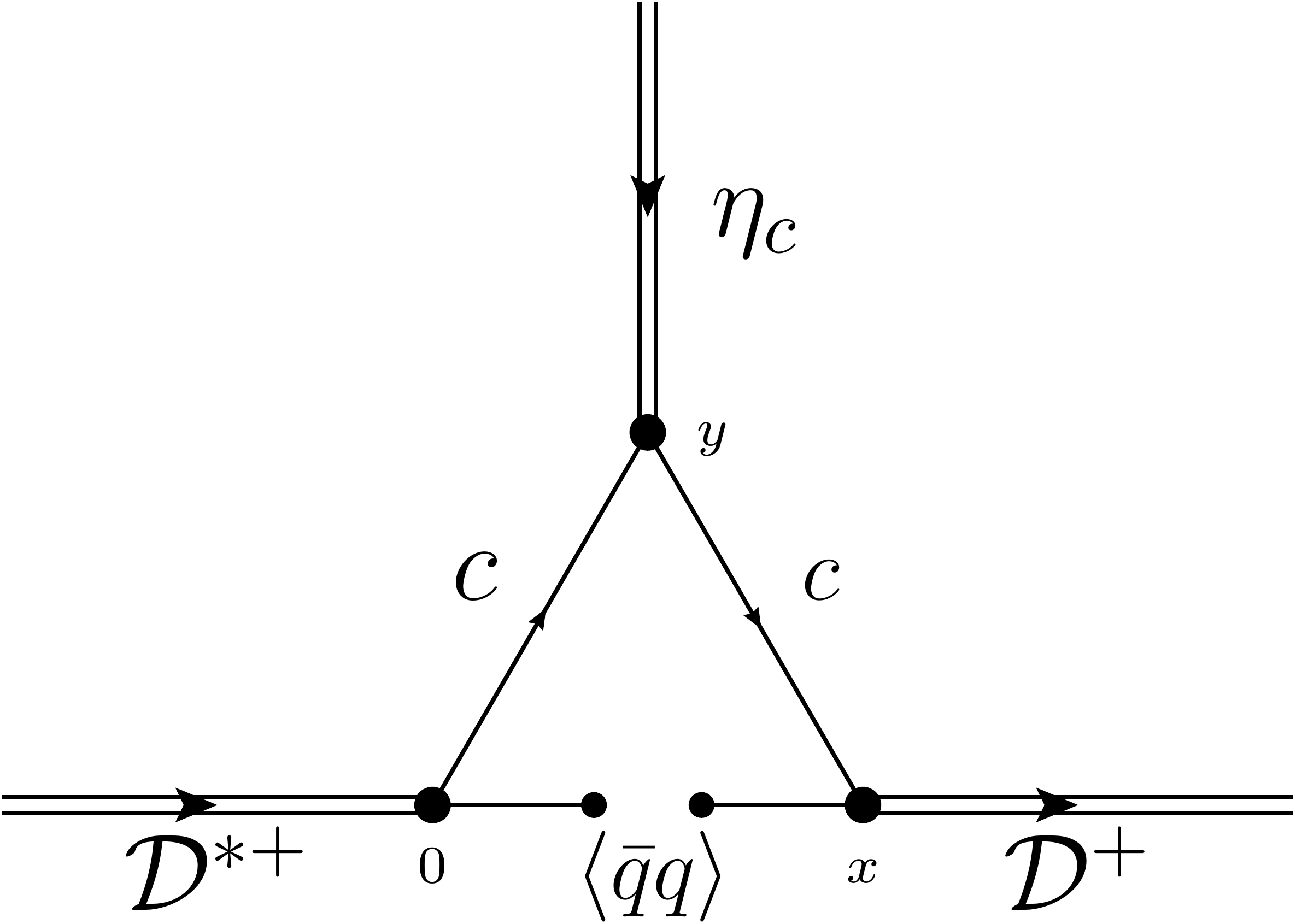}\label{subfig:b}}\quad
 \subfigure[]{\includegraphics[width=0.29\linewidth]{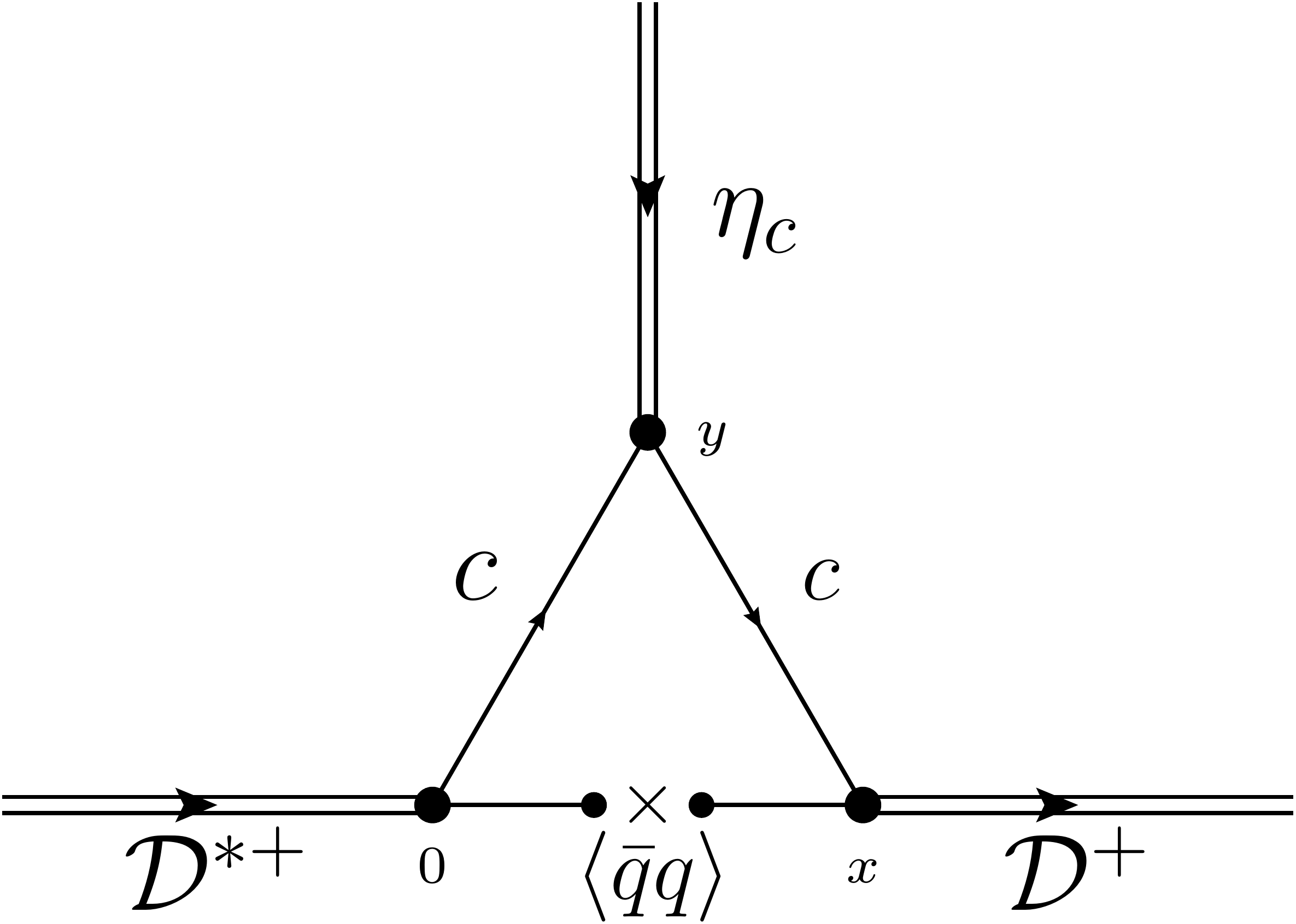}\label{subfig:c}}\\
 \subfigure[]{\includegraphics[width=0.29\linewidth]{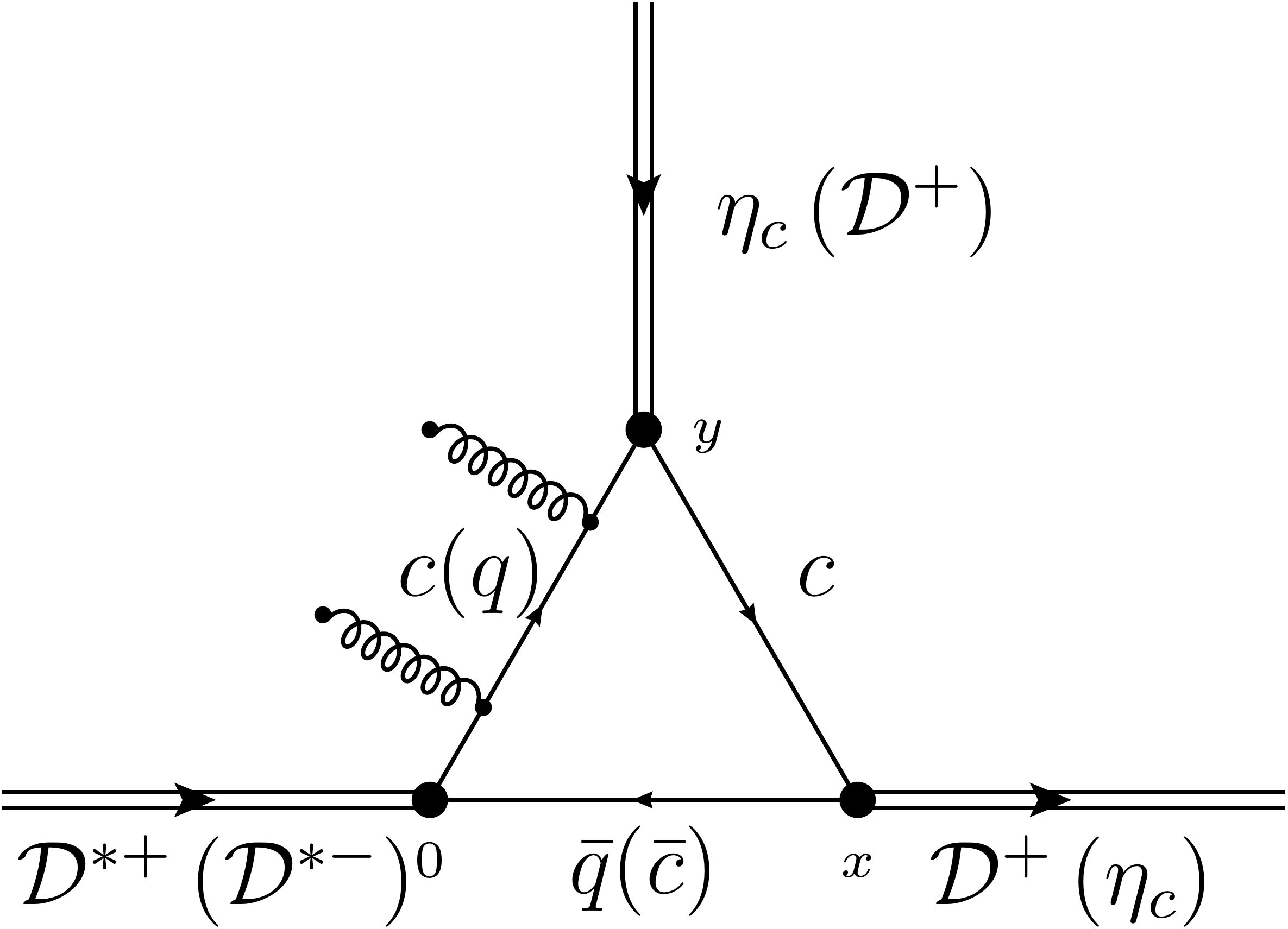}\label{subfig:d}}\quad
 \subfigure[]{\includegraphics[width=0.29\linewidth]{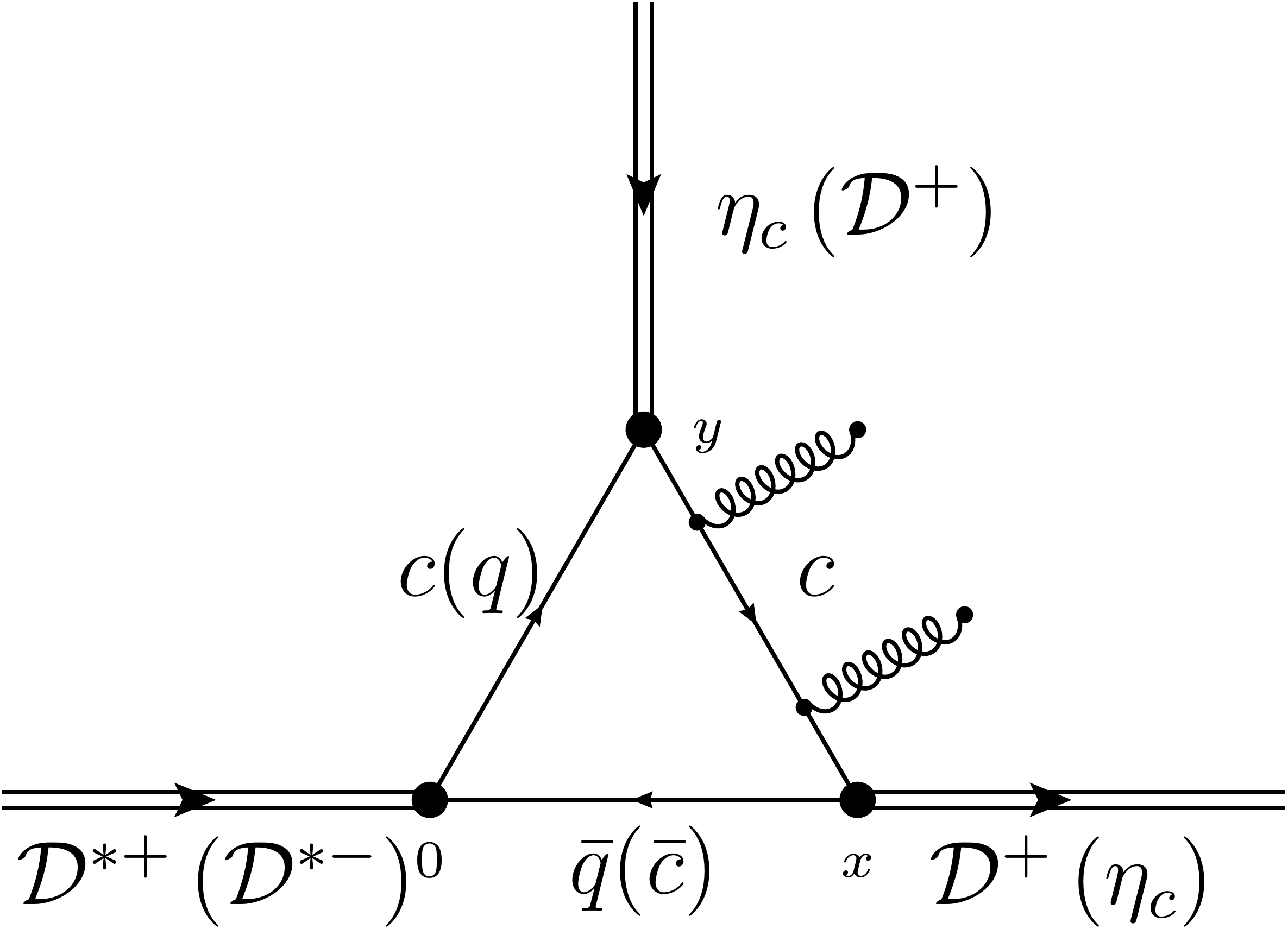}\label{subfig:e}}\quad
 \subfigure[]{\includegraphics[width=0.29\linewidth]{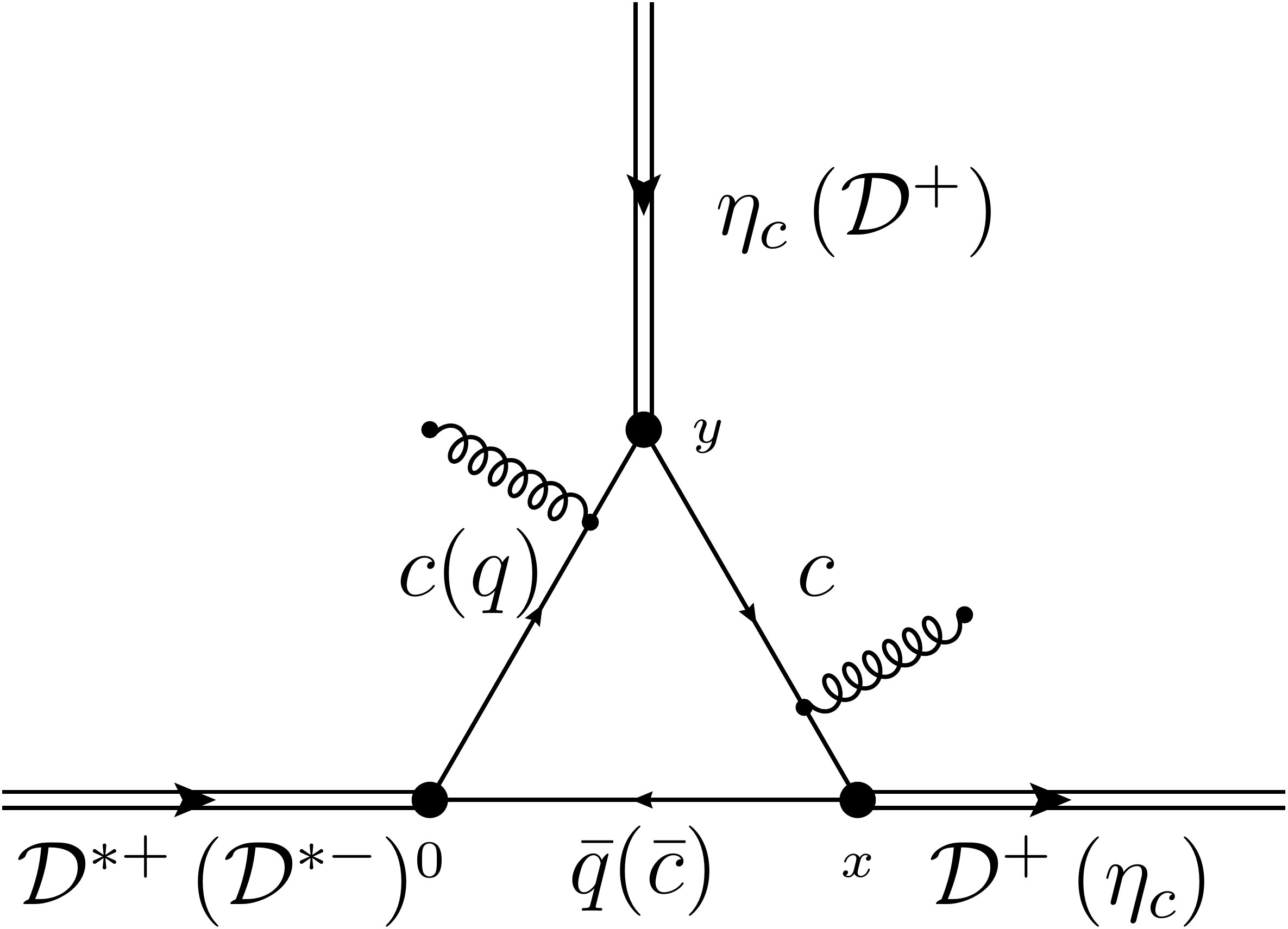}\label{subfig:f}} \\
 \subfigure[]{\includegraphics[width=0.29\linewidth]{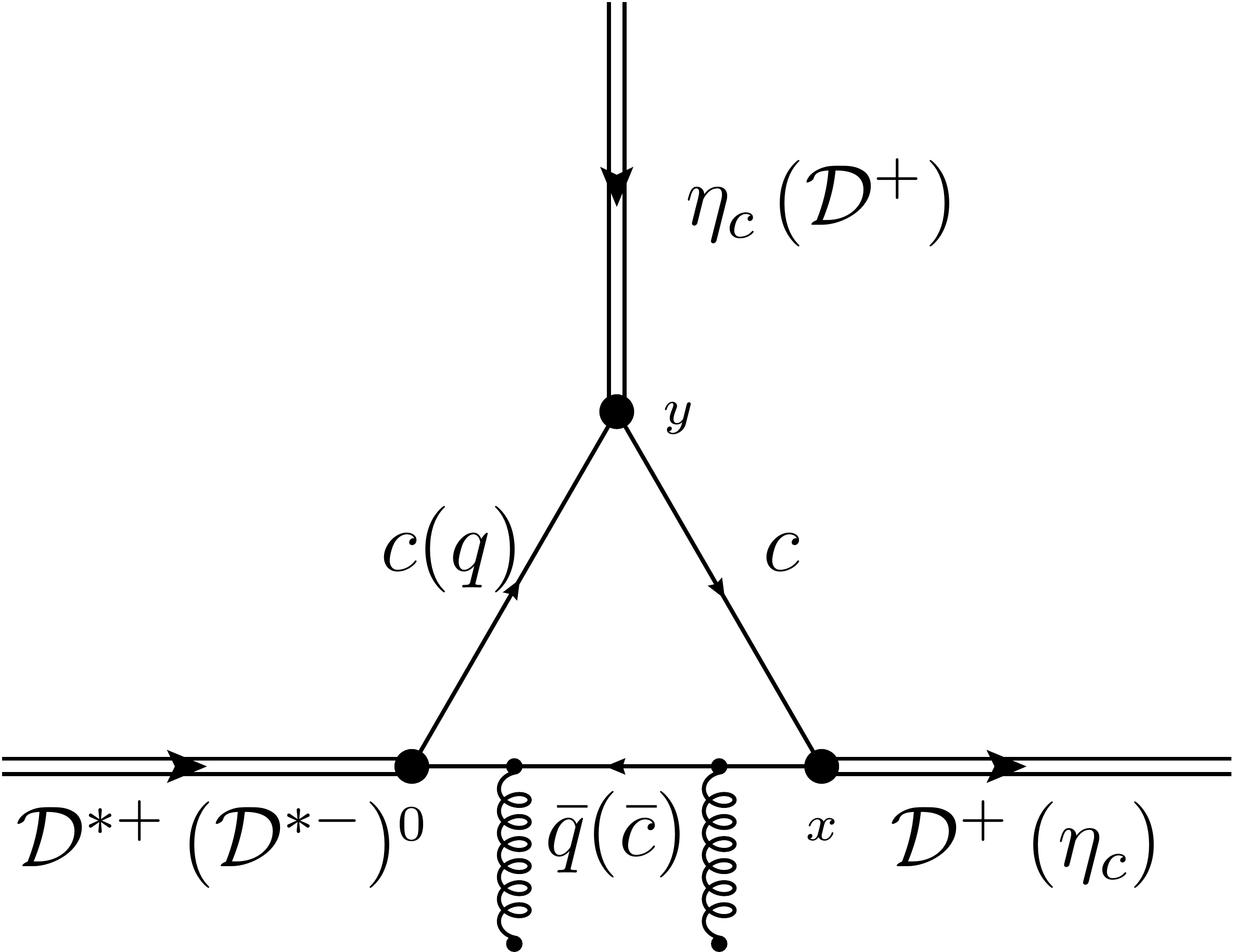}\label{subfig:g}}\quad
 \subfigure[]{\includegraphics[width=0.29\linewidth]{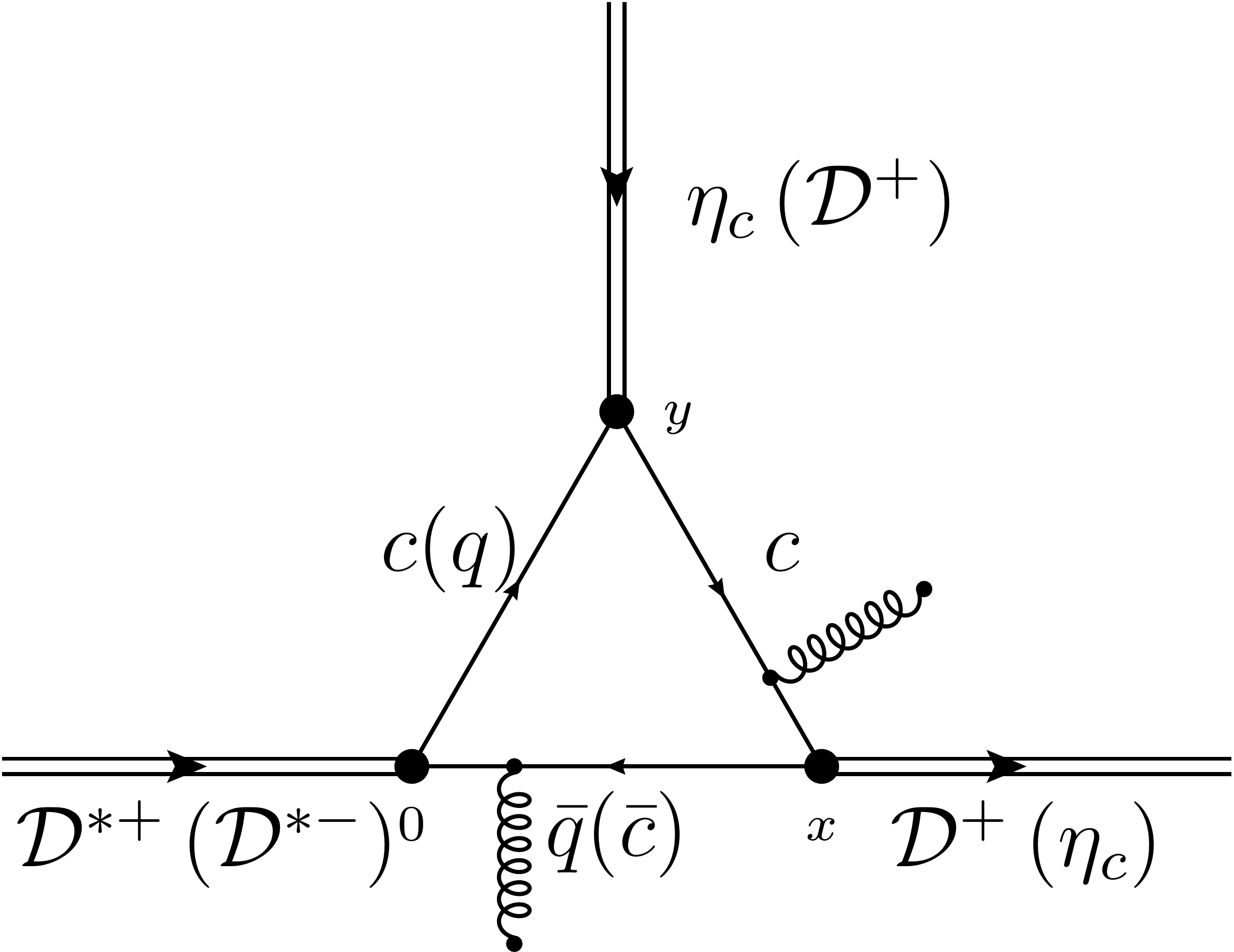}\label{subfig:h}}\quad
 \subfigure[]{\includegraphics[width=0.29\linewidth]{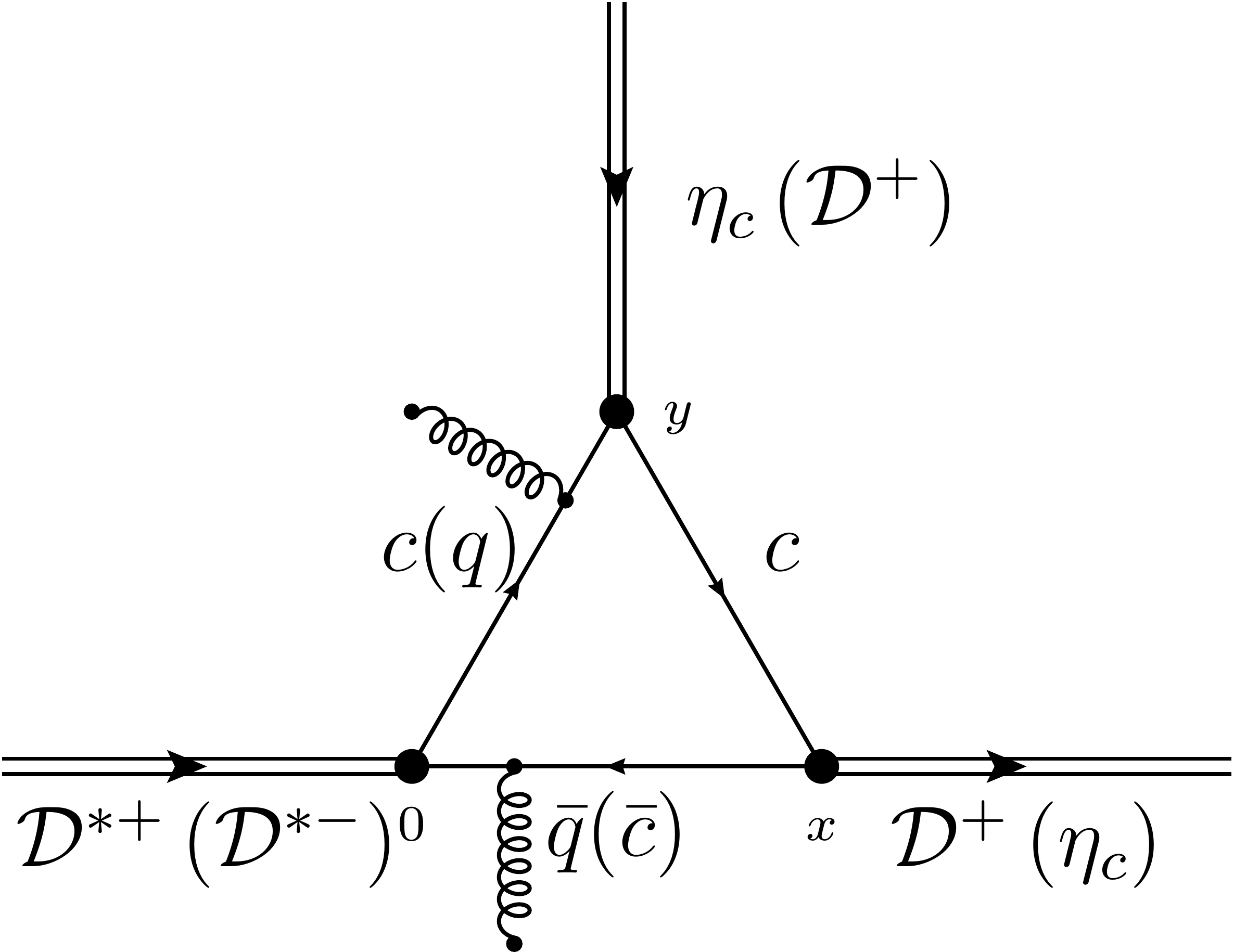}\label{subfig:i}} \\
 \subfigure[]{\includegraphics[width=0.29\linewidth]{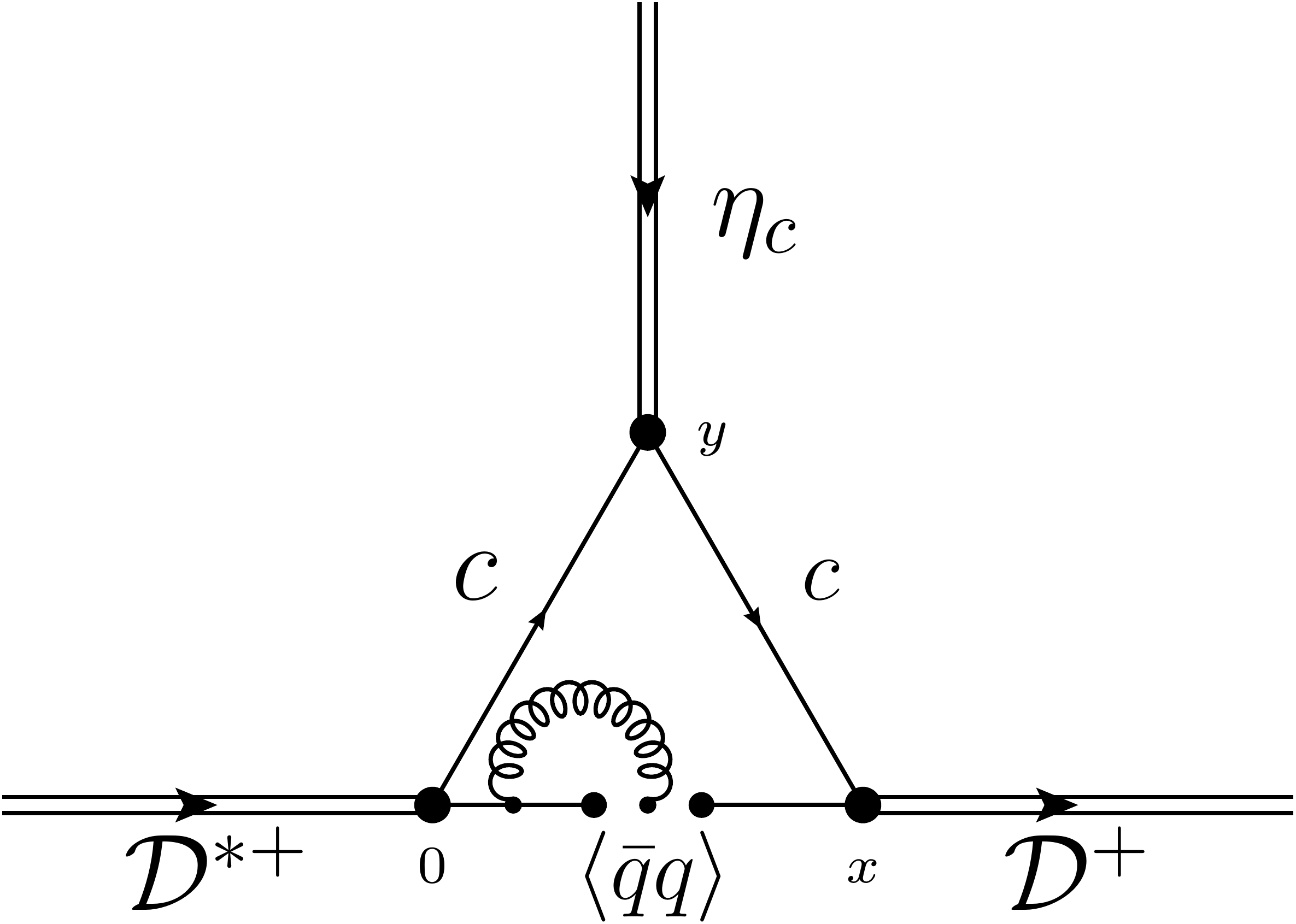}\label{subfig:j}}\quad
 \subfigure[]{\includegraphics[width=0.29\linewidth]{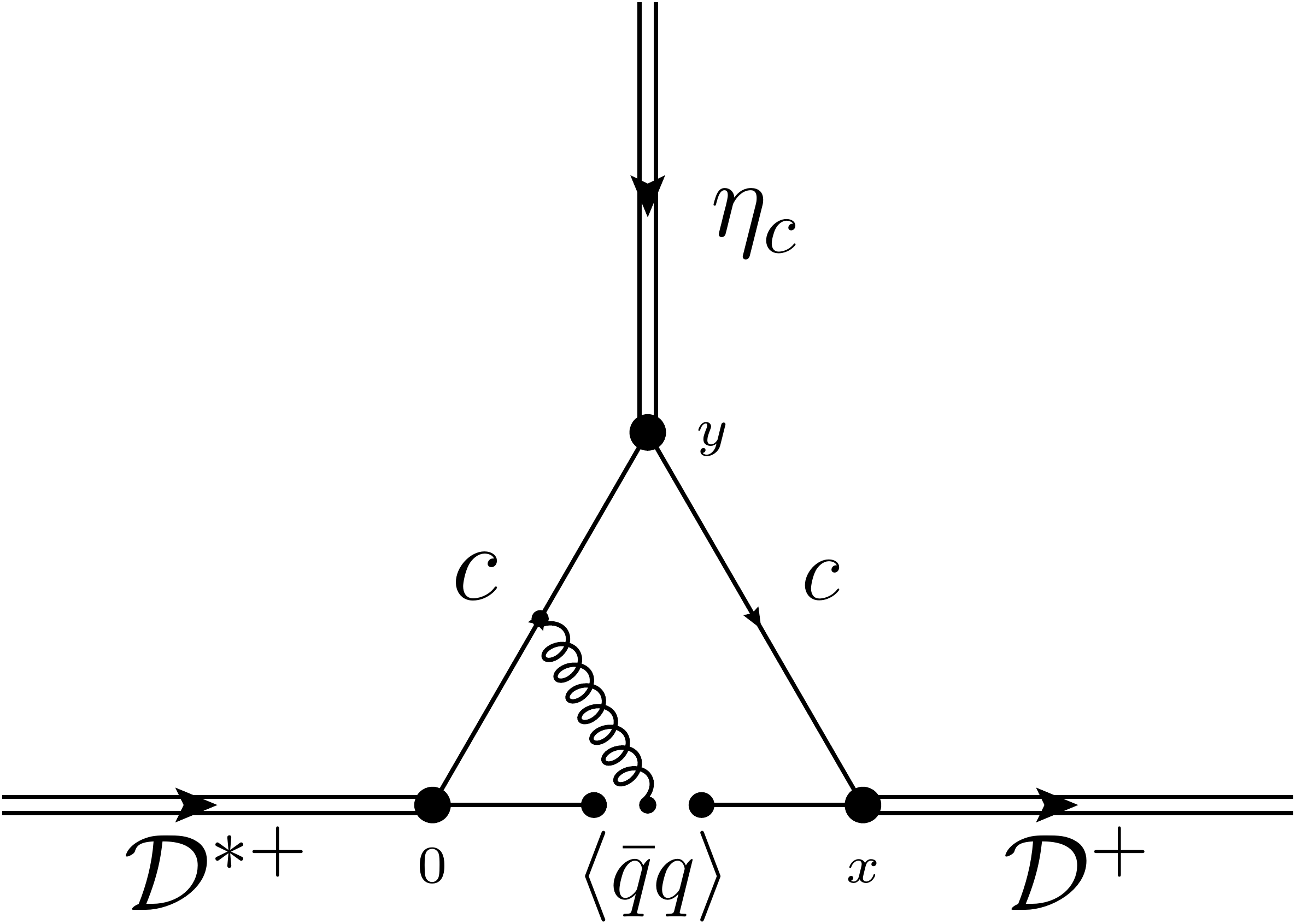}\label{subfig:k}} \quad
 \subfigure[]{\includegraphics[width=0.29\linewidth]{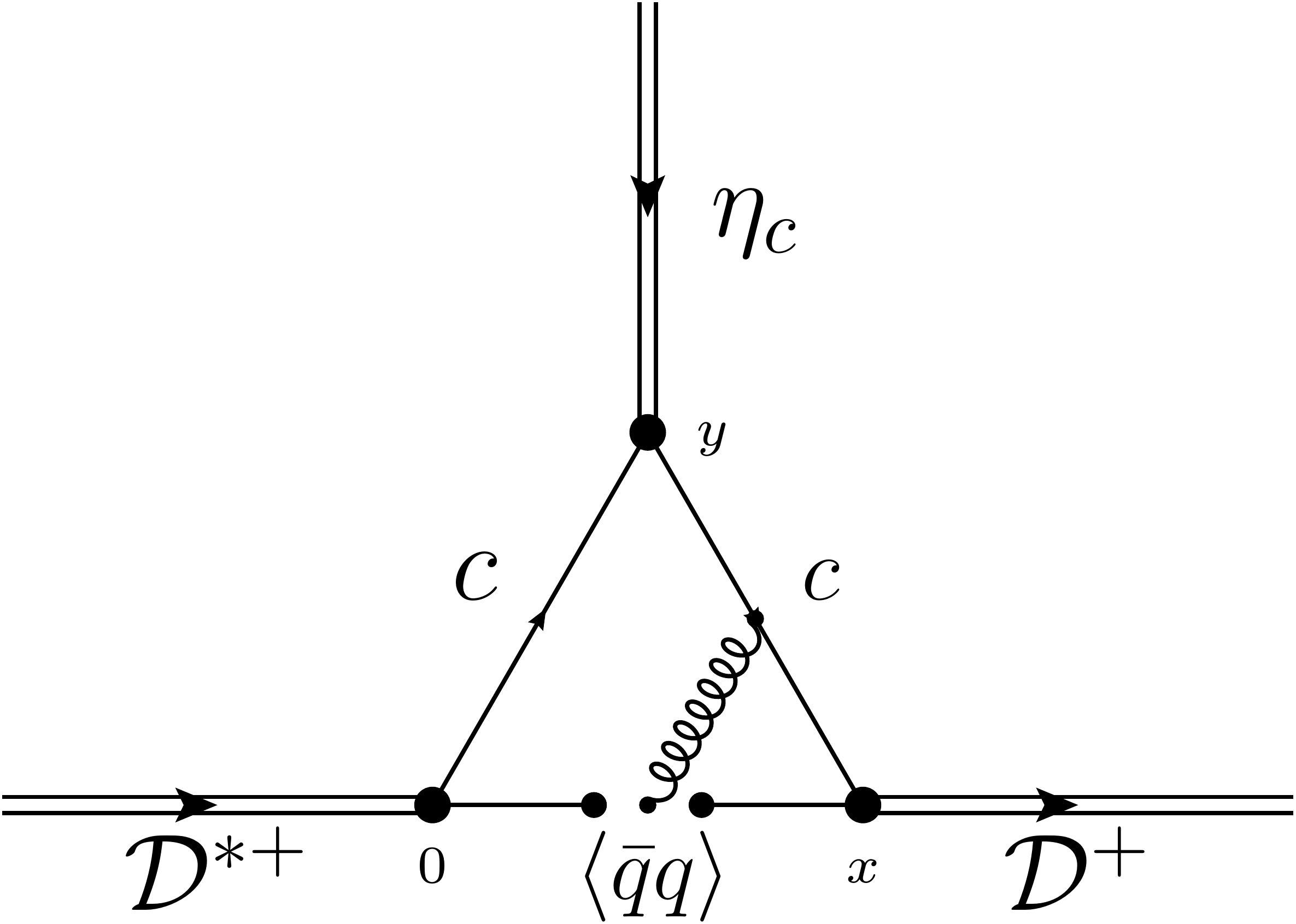}\label{subfig:l}} \\
 \subfigure[]{\includegraphics[width=0.29\linewidth]{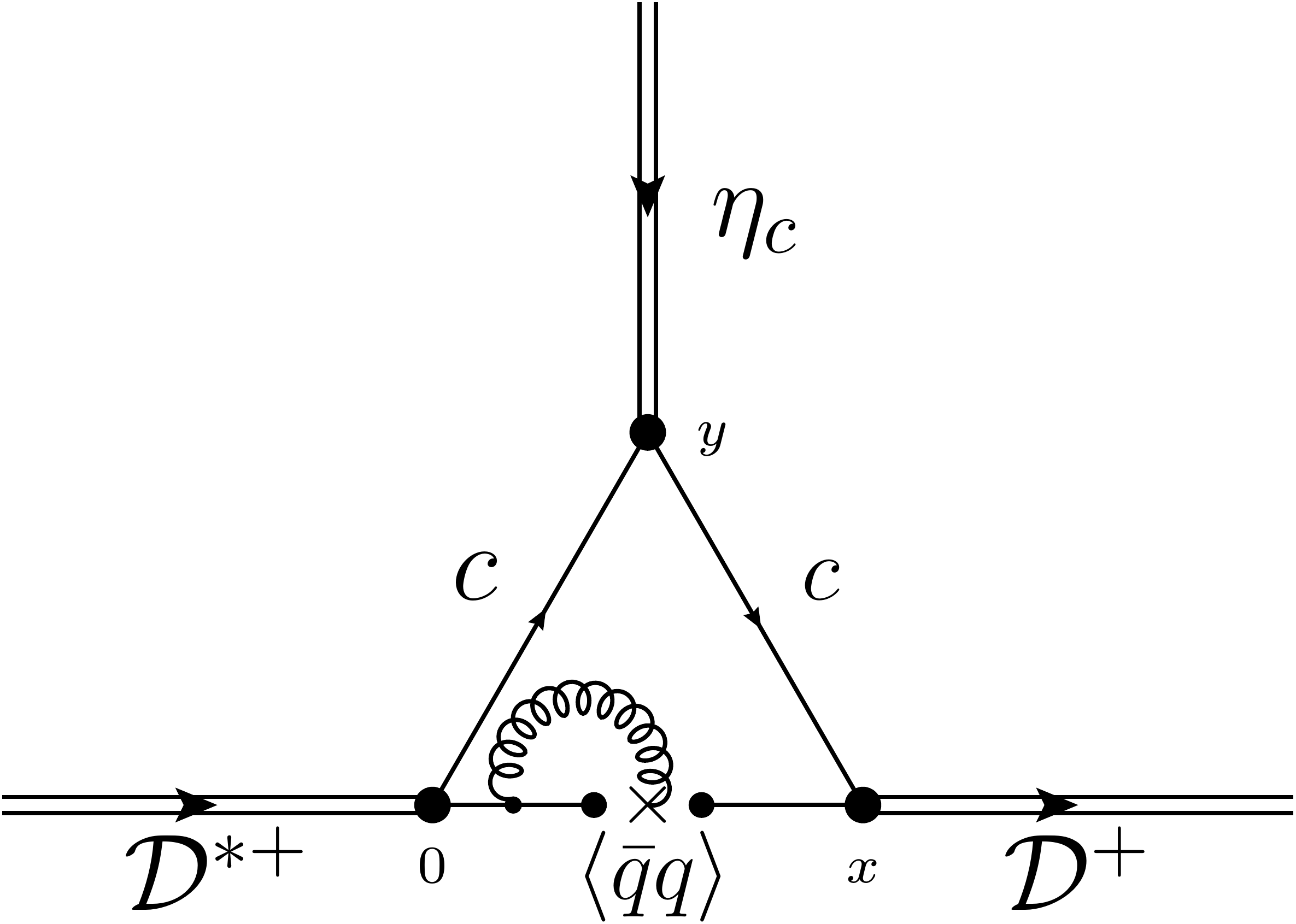}\label{subfig:m}} \quad
 \subfigure[]{\includegraphics[width=0.29\linewidth]{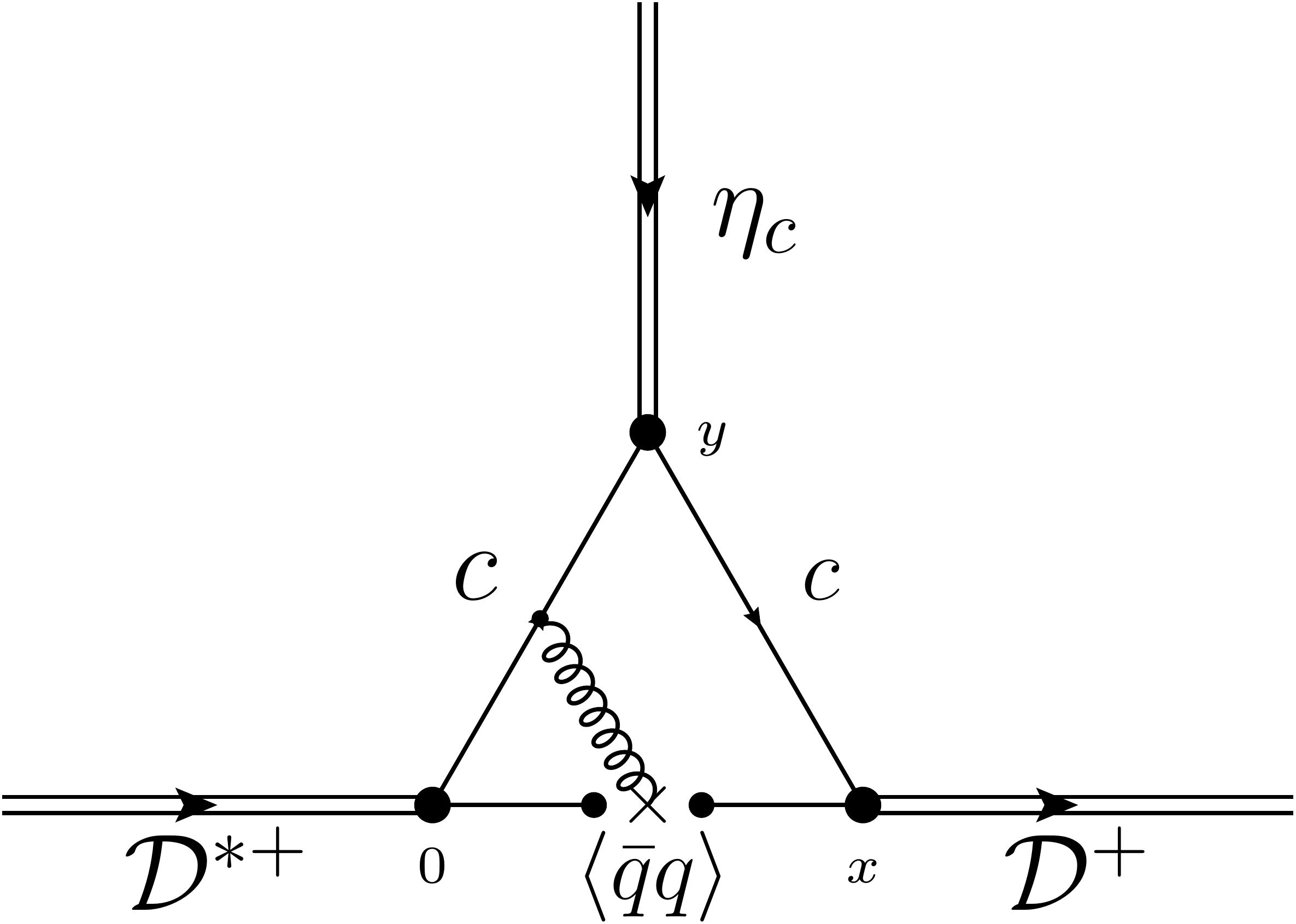}\label{subfig:n}}\quad
 \subfigure[]{\includegraphics[width=0.29\linewidth]{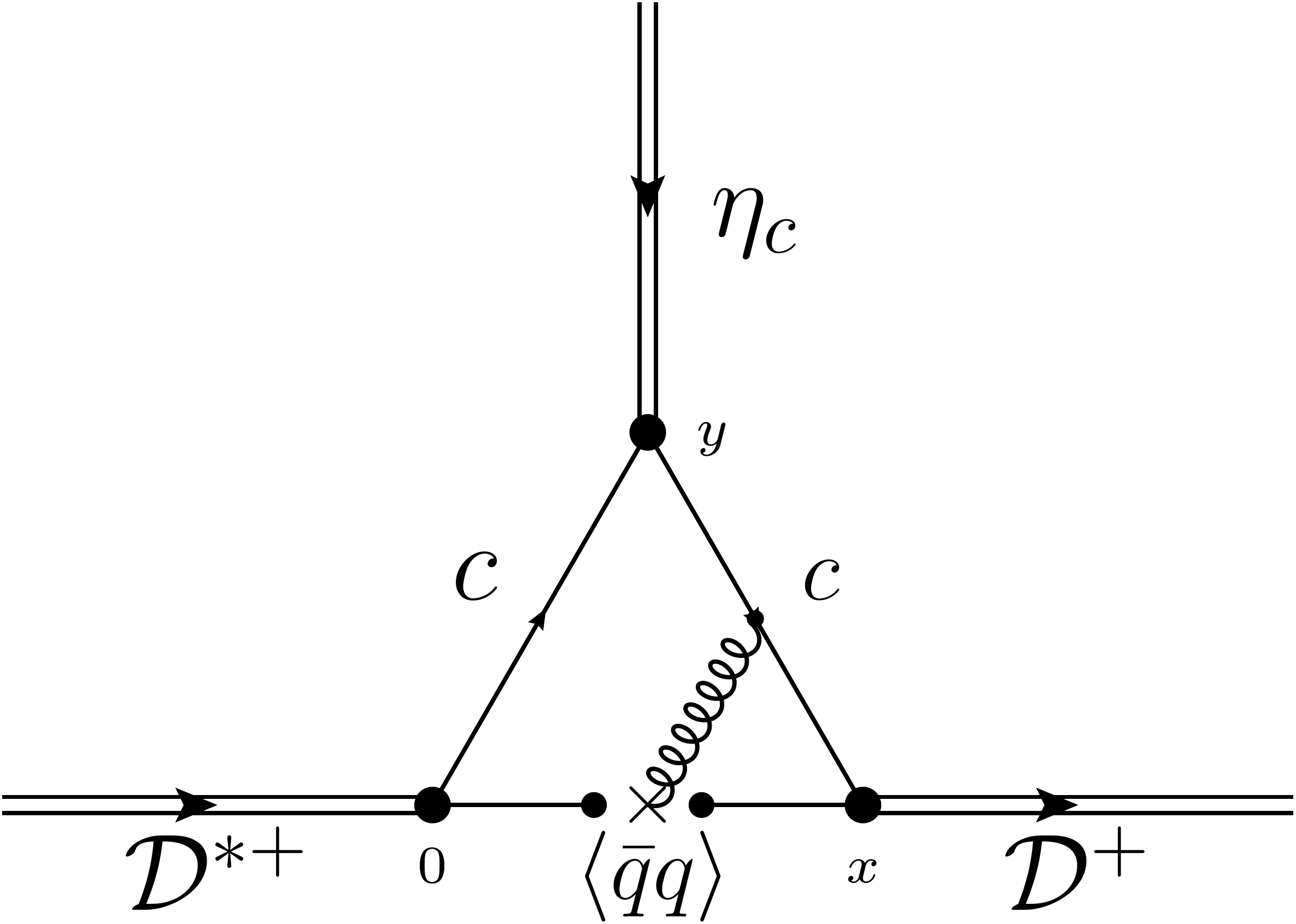}\label{subfig:o}}
\caption{\label{fig:diagrams}Contributing OPE diagrams for $\eta_c (\mathcal{D})$ off-shell. }
\end{figure}

The  diagrams  contributing  to  the correlation  functions  that  are
calculated in this work  are shown in Fig.~\ref{fig:diagrams}.  Notice
that there  are suppressed diagrams  that are omitted in  this figure.
The case of off-shell $\eta_c$ is  the only one that has contributions
from      all     the      non-perturbative      terms     of      the
Eq.~(\ref{eq:nonpertcontrib}),  this being  an  effect  of the  double
Borel transformation that suppresses all the non-perturbative diagrams
except  for the  gluon condensates  (Fig.~\ref{fig:diagrams}(d-i)) for
the cases of off-shell $\mathcal{D}$ and $\mathcal{D}^*$.

The   perturbative   term   for    a   given   off-shell   meson   $M$
(Fig.~\ref{fig:diagrams}(a)) can  be written in terms  of a dispersion
relation:
\begin{align}
\Gamma^{\text{pert} (M)}_\mu(p,p') =  - \frac{1}{4\pi^2} \int^\infty_0
\int^\infty_0       \frac{\rho_\mu^{\text{pert}       (M)}(s,       u,
  t)}{(s-p^2)(u-p'^2)} ds du\,.
\label{eq:pertgeral}
\end{align}
where $\rho_\mu^{\text{pert} (M)}(s, u, t)$ is the spectral density of
the perturbative term,  which is related to the imaginary  part of the
correlation              function,              $\rho_\mu^{\text{pert}
  (M)}(s,u,t)=\frac{1}{\pi}{\text{Im}}[\Gamma^{\text{pert}(M)}_\mu(s,u,t)]$.            The
following expression  is  obtained by applying the Cutkosky
rules and by the use of Lorentz symmetries:
\begin{align}
\rho_\mu^{\text{pert}  (M)}(s,u,t) =  \frac{3}{2\sqrt{\lambda}}\left [
  F_p^{(M)}(s,u,t)p_\mu + F_{p'}^{(M)}(s,u,t)p'_\mu \right ] \,,
\label{dspec}
\end{align}
where $\lambda = (u  + s - t)^2 - 4us$,  and the functions $F_p^{(M)}$
and $F_{p'}^{(M)}$ are the invariant amplitudes. For the cases studied
in this work, these amplitudes can be written as:
\begin{eqnarray}
F_p^{(\eta_c)} &=& A(s-u-t) - u - (m_q-m_c)^2\,,\\
F_{p'}^{(\eta_c)} &=& B(s-u-t) - s + (m_q-m_c)^2\,,\\
F_p^{(\mathcal{D}^*)} &=& A(s+u-t) - u  \,,\\
F_{p'}^{(\mathcal{D}^*)} &=& B(s+u-t) - s + (m_q - m_c)^2\,,\\
F_p^{(\mathcal{D})} &=& A(s-u-t) + u \,,\\
F_{p'}^{(\mathcal{D})} &=& B(s-u-t) - s + (m_q - m_c)^2\,,
\end{eqnarray}
where
\begin{align}
&A = \left [ \frac{\bar{k}_0}{\sqrt{s}} - \frac{p'_0 \overline{|\vec{k}|} \overline{\cos\theta}}{|\vec{p'}|\sqrt{s}} \right ]\,,
\;\;\;\;\;\;\;\;
B = \frac{\overline{|\vec{k}|} \overline{\cos\theta}}{|\vec{p'}|} \,,
\;\;\;\;\;\;\;\;
\bar{k}_0 = \frac{s + \epsilon(m_q^2-m_c^2)}{2\sqrt{s}}\,,
\nonumber \\
&\overline{|\vec{k}|} = \sqrt{\bar{k}_0^2 + \frac{(\epsilon-1)}{2} m_c^2 - \frac{(\epsilon+1)}{2} m_q^2}\,,
\;\;\;\;\;\;\;\;
\overline{\cos\theta} = \frac{2p'_0\bar{k}_0 - u + \frac{1+\epsilon}{2}(m_c^2 - m_q^2) }{2|\vec{p'}|\overline{|\vec{k}|}} \,,
\nonumber \\
&p'_0 = \frac{s+u-t}{2\sqrt{s}}\,,\;\;\;\;\;\;\;\;|\vec{p'}| = \frac{\sqrt{\lambda}}{2\sqrt{s}}\,, \nonumber
\end{align}
and      $\epsilon      =      1(-1)$      for      the      off-shell
$\eta_c(\mathcal{D}^{(*)})$.      The     quantities      $\bar{k}_0$,
$\overline{|\vec{k}|}$, and $\overline{\cos\theta}$  are the centers
of  the  $\delta$-functions that are present in the Cutkosky rules, and
$s=p^2$, $u=p'^2$, $t=q^2$ are the Mandelstam variables.

The  first  non-perturbative  terms contributing  to  the  correlation
function is  the quark  condensate $\langle\bar{q}q\rangle$,  shown on
the (b)-diagram of Fig.~\ref{fig:diagrams}:
\begin{align}
\Gamma^{\langle\bar{q}q\rangle (\eta_c)}_\mu = \frac{m_c \langle\bar{q}q\rangle[p_\mu + p'_\mu ]}{(p^2 - m^2_c)(p'^2 - m_c^2)}\,.
\label{eq:contribcond}
\end{align}
The (c)- diagram shown  on Fig.~\ref{fig:diagrams} represents the mass
term of  the quark  condensate, $m_q\langle\bar{q}q\rangle$,  which is
numerically  suppressed due  to  the  low values  of  the light  quark
masses $m_q$:
\begin{align}
\Gamma^{m_q\langle\bar{q}q\rangle (\eta_c)}_\mu = m_q\langle\bar{q}q\rangle\frac{ 
\left [ 2m_c^2( p^2-p'\cdot p -\frac{m_c^2}{2}) + p'^2(2p'\cdot p - p^2)\right ]p_\mu + m_c^2(2m_c^2 - p^2-p'^2)p'_\mu
}{2\left (p^2 - m_c^2 \right )^2 \left ({p'}^2 - m_c^2 \right )^2}\,.
\end{align}

Contributions from the  charm quark condensate are very  small and can
be safely  neglected. The  complete expressions for  the contributions
from    gluon     condensates    ($\langle    g^2     G^2    \rangle$,
Fig.~\ref{fig:diagrams}(d-i)), and from quark-gluons mixed condensates
($\langle      \bar{q}g\sigma      \cdot       G      q      \rangle$,
Fig.~\ref{fig:diagrams}(j-o)) can be  found in the Appendix  A for the
case of the off-shell $\eta_c$.

\subsection{The sum rule}

The sum rule is obtained using the quark-hadron duality principle,
matching the phenomenological and OPE sides:
\begin{equation}
{\cal BB}\left[\Gamma_\mu^{OPE(M)}\right](M,M')={\cal BB}\left[\Gamma_\mu^{phen(M)}\right](M,M')\,, \label{qhb}
\end{equation}
where    the    double    Borel    transform    ($\cal    BB$)    was
applied~\cite{Khodjamirian:2002pka,Colangelo:2000dp},     with     the
following variable transformations: $P^2 = -  p^2 \to M^2$ and $P'^2 =
-p'^2 \to M'^2$, where $M$ and $M'$ are the Borel masses.

In   order  to   eliminate   the  $h.r.$   terms   appearing  on   the
phenomenological               side               of               the
Eqs.~(\ref{eq:fenomEtacoff})-(\ref{eq:fenomDoff}),    the    threshold
continuum parameters, $s_0$ and $u_0$, are introduced in the limits of
the  integrals on  the  OPE  side. These  are  cutoff parameters  that
satisfy the  relations $m_i^2  < s_0  < {m'}_i^2$ and  $m_o^2 <  u_0 <
{m'}_o^2$, where $m_i$ and $m_o$  are, respectively, the masses of the
off-shell  mesons that  comes  in and  out of  the  diagrams shown  on
Fig.~\ref{fig:diagrams}, and  $m'$ is  the mass  of the  first excited
state  of such  mesons. The  application of  the quark-hadron  duality
principle  allows us  to identify  that the  integrals from  $s_0$ and
$u_0$ up to infinity in the Eq.~(\ref{eq:pertgeral}) correspond to the
$h.r.$ terms on  the phenomenological side, thus  canceling such terms
from the sum rules.

After these two steps, the Eq.~(\ref{qhb})  can be used to compute the
expressions for  the form  factors of  each one  of the  off-shell mass
cases. The  expressions for the  structures $p_\mu$ of the  cases with
off-shell $\eta_c$ and  $\mathcal{D}^*$, and $p'_\mu$ for  the case with
off-shell $\mathcal{D}$ are given by:
\begin{multline}
g_{\eta_c     \mathcal{D}^*     \mathcal{D}}^{(\eta_c    )}(Q^2)     =
\frac{-\frac{3}{8\pi^2}    \int^{s_0}_{s_{inf}}   \int^{u_0}_{u_{inf}}
  \frac{1}{\sqrt{\lambda}}F_{p_\mu}^{(\eta_c)}
  e^{-\frac{s}{M^2}}e^{-\frac{u}{M'^2}}         ds        du         +
  \mathcal{B}\mathcal{B}\left   [  \Gamma^{\text{non--pert}}_{{p_\mu}}
    \right]}{\frac{-C(Q^2+m^2_{\mathcal{D}}                          +
    m^2_{\mathcal{D}^*})}{(Q^2+m_{\eta_c}^2)}e^{-m_{\mathcal{D}^*}^2/M^2}e^{-m_{\mathcal{D}}^2/M'^2}}\,,\label{eq:rsqcdetac}
\end{multline}
\begin{multline}
g_{\eta_c   \mathcal{D}^*   \mathcal{D}}^{(\mathcal{D}^*   )}(Q^2)   =
\frac{-\frac{3}{8\pi^2}    \int^{s_0}_{s_{inf}}   \int^{u_0}_{u_{inf}}
  \frac{1}{\sqrt{\lambda}}F_{p_\mu}^{(\mathcal{D}^*)}
  e^{-\frac{s}{M^2}}e^{-\frac{u}{M'^2}}         ds        du         +
  \mathcal{B}\mathcal{B}\left       [        \Gamma^{\langle       g^2
      G^2\rangle}_{{p_\mu}}
    \right]}{\frac{C(m^2_{\eta_c}-m^2_{\mathcal{D}}                  +
    m^2_{\mathcal{D}^*})}{(Q^2+m_{\mathcal{D}^*}^2)}e^{-m_{\mathcal{D}}^2/M^2}e^{-m_{\eta_c}^2/M'^2}}\,,\label{eq:rsqcdDEst}
\end{multline}
\begin{multline}
g_{\eta_c    \mathcal{D}^*   \mathcal{D}}^{(\mathcal{D}    )}(Q^2)   =
\frac{-\frac{3}{8\pi^2}    \int^{s_0}_{s_{inf}}   \int^{u_0}_{u_{inf}}
  \frac{1}{\sqrt{\lambda}}F_{p'_\mu}^{(\mathcal{D})}
  e^{-\frac{s}{M^2}}e^{-\frac{u}{M'^2}}         ds        du         +
  \mathcal{B}\mathcal{B}\left     [     \Gamma^{\langle    g^2     G^2
      \rangle}_{{p'_\mu}}              \right]}{\frac{2              C
    m^2_{\mathcal{D}^*}}{(Q^2+m_{\mathcal{D}}^2)}e^{-m_{\mathcal{D}^*}^2/M^2}e^{-m_{\eta_c}^2/M'^2}}\,,\label{eq:rsqcdD}
\end{multline}
where the constant $C$ is defined in the Eq.~(\ref{eq:C}).

The coupling constant $g_{\eta_c \mathcal{D}^*
  \mathcal{D}}$ is defined as:
\begin{align}
g_{\eta_c \mathcal{D}^* \mathcal{D}} = \lim_{Q^2\to -m^2_M} g_{\eta_c \mathcal{D}^* \mathcal{D}}^{(M)}(Q^2)
\end{align}
where $M$  is once again  the off-shell meson.

The above  expression implies  that, in order  to obtain  the coupling
constant, it is necessary to  extrapolate the numerical results of the
form factors  to the region $Q^2  < 0$, outside of  the deep Euclidean
region     where     the     QCDSR    are     valid.      From     the
Eqs.~(\ref{eq:rsqcdetac}),~(\ref{eq:rsqcdDEst}) and~(\ref{eq:rsqcdD}),
it is  clear that  it is  possible to  evaluate the   coupling constant
$g_{\eta_c  \mathcal{D}^*  \mathcal{D}}$   from  three  distinct  form
factors, one for each case  of off-shell mass. However, these coupling
constants must present the same  value, regardless of the extrapolated
form factor.   This condition  is used  to minimize  the uncertainties
existing in the coupling constant calculation,  as it will be clear in
the next section.

\section{Results and Discussion}
The      Eqs.~(\ref{eq:rsqcdetac}),      (\ref{eq:rsqcdDEst})      and
(\ref{eq:rsqcdD}) shows the  three different form factors  that can be
obtained for the vertices. In order to minimize the uncertainties that
comes from the  extrapolation of the QCD results, it  is required that
the three  form factors  lead to  the same  coupling constant  for the
limits    $Q^2    =    -M^2$     ($M    =    \eta_c,    \mathcal{D}^*,
\mathcal{D}$)~\cite{Bracco:2001pqq}.    The  Table~\ref{tab:mesonmass}
shows the values  of the masses of  the mesons that were  used in this
work.

\begin{table}[ht]
\caption{\label{tab:mesonmass}Masses of the mesons present in this work.}

\begin{tabular*}{0.8\linewidth}{@{\extracolsep{\fill}} cccccc}
\hline
Meson & $\eta_c$ & $D^*$ & $D$ & $D_s^*$ & $D_s$ \\
\hline
Mass (GeV) \cite{Agashe:2014kda} & 2.983 & 2.010 & 1.869 & 2.112 & 1.968 \\
%$f$ (MeV) & $394.7 \pm 2.4$  & $242^{+20}_{-12}$ & $206.7 \pm 11.0$ &$301\pm 13$  & $257.5 \pm 6.1$ & -- & -- & --\\
%$\langle \bar{q}q\rangle$ (MeV$^3$) & -- & -- & --& --& -- & $-(290\pm 15)^3$ & $-(290\pm 15)^3$& 0\\
%$\langle \bar{q}g\sigma \cdot G q \rangle$ (GeV$^5$) & -- & -- & --& --& -- & & & 0\\
\hline
\end{tabular*}
\end{table}

We consider that reliable results are  obtained from the QCDSR if such
results show  good stability in relation  to the Borel masses  $M$ and
$M'$. The set  of values of the  Borel masses for which  the QCDSR are
stable  are called  the  "Borel  window".  The  window  is defined  by
imposing that the pole contribution  must be larger than the continuum
contribution and that the contribution of the perturbative term of the
OPE must be at least $50\%$ of the total.

We use the  ansatz $M'^2 = \frac{m_o^2}{m_i^2}M^2$,  which relates the
Borel masses $M$ and $M'$, decreasing the computational effort for the
QCDSR calculations. In such ansatz, $m_o$  and $m_i$ are the masses of
the   mesons   related   with   the   quadrimomenta   $p'$   and   $p$
respectively. For  each value  of $Q^2$,  the mean  value of  the form
factors  are computed  within the  Borel window,  and thus  it is  not
necessary to  choose to work  with a fixed  value for the  Borel mass,
minimizing     the     uncertainties      associated     with     this
parameter~\cite{OsorioRodrigues2014143,JpsiDsEstDs}.    The   standard
deviation is  then used to automate  the analysis of the  stability of
the form factors related to the Borel mass and the continuum threshold
parameters.  The  criterion of the Borel  window stability establishes
the optimal values  that must be used for the  continuum threshold and
the Borel window parameters. Therefore, not only the stability will be
assured  within the  Borel window,  but  also within  the interval  of
values of $Q^2$ that will be used.

\begin{figure}[!ht]
\centering
\subfigure[]{\includegraphics[width=0.49\linewidth]{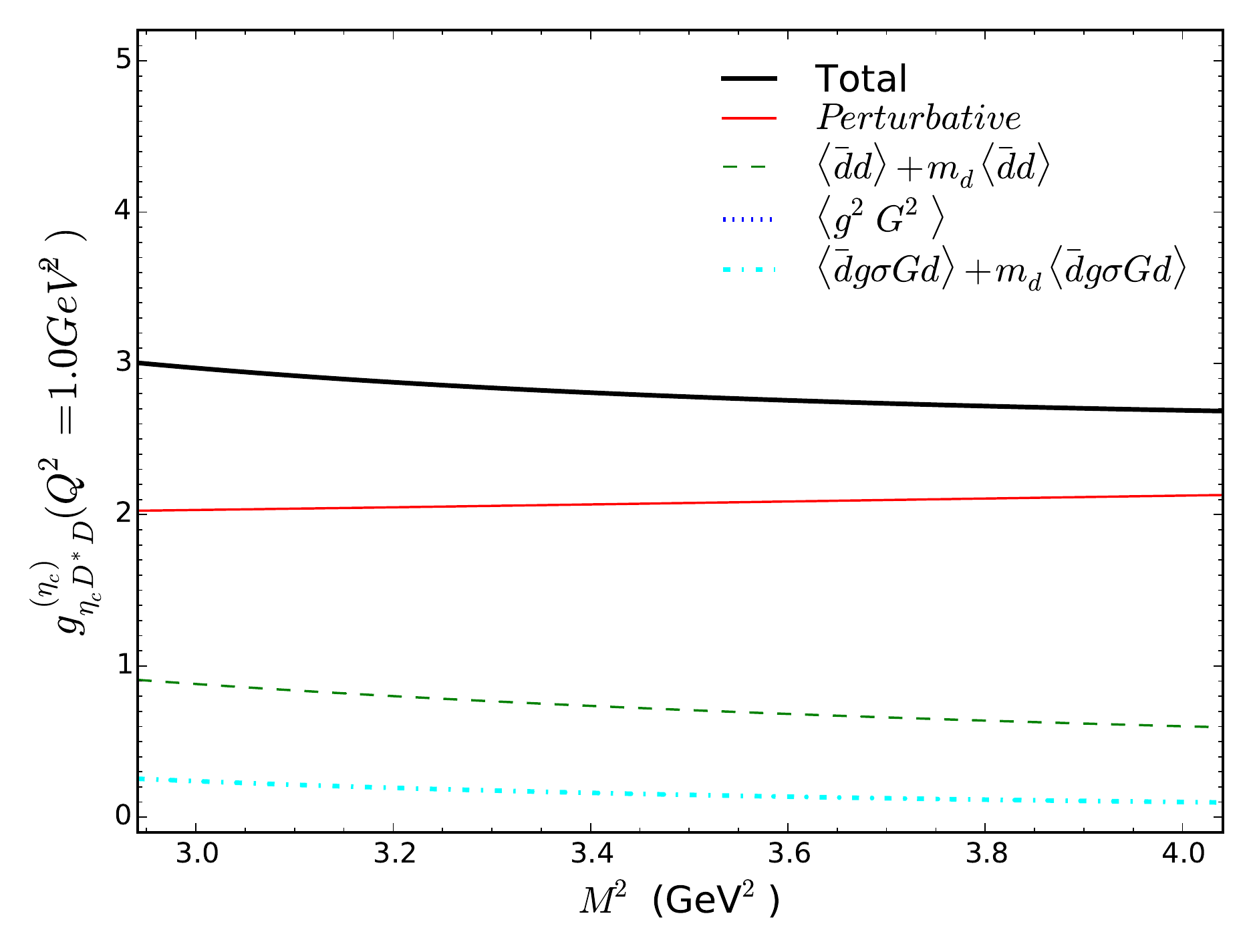}\label{subfigestabilidade:a}}
\subfigure[]{\includegraphics[width=0.49\linewidth]{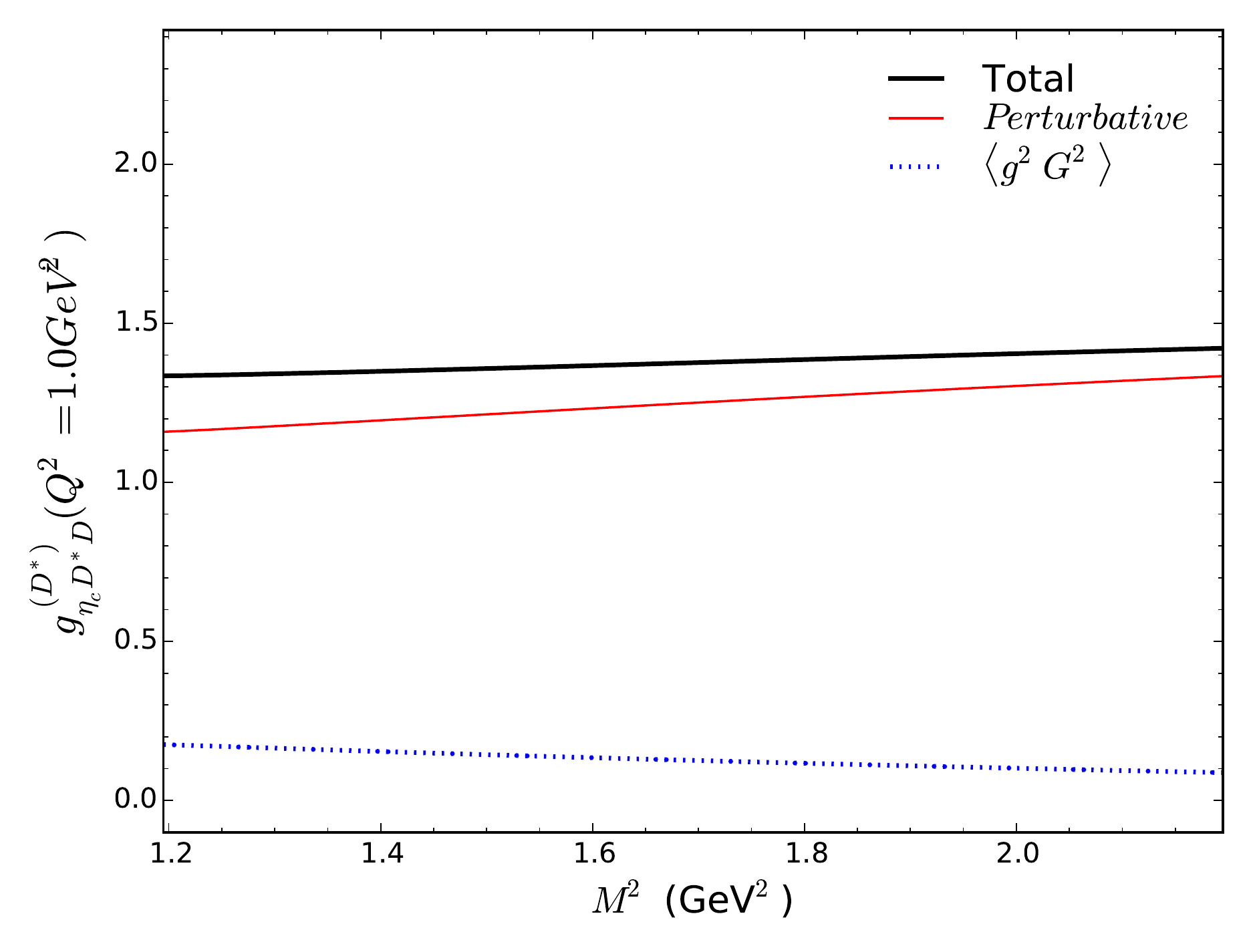}\label{subfigestabilidade:b}}\\
	%\subfigure[]{\includegraphics[width=0.49\linewidth]{figs/EtacDEstD/Contrib_Dpl.pdf}\label{subfigestabilidade:c}}
\subfigure[]{\includegraphics[width=0.49\linewidth]{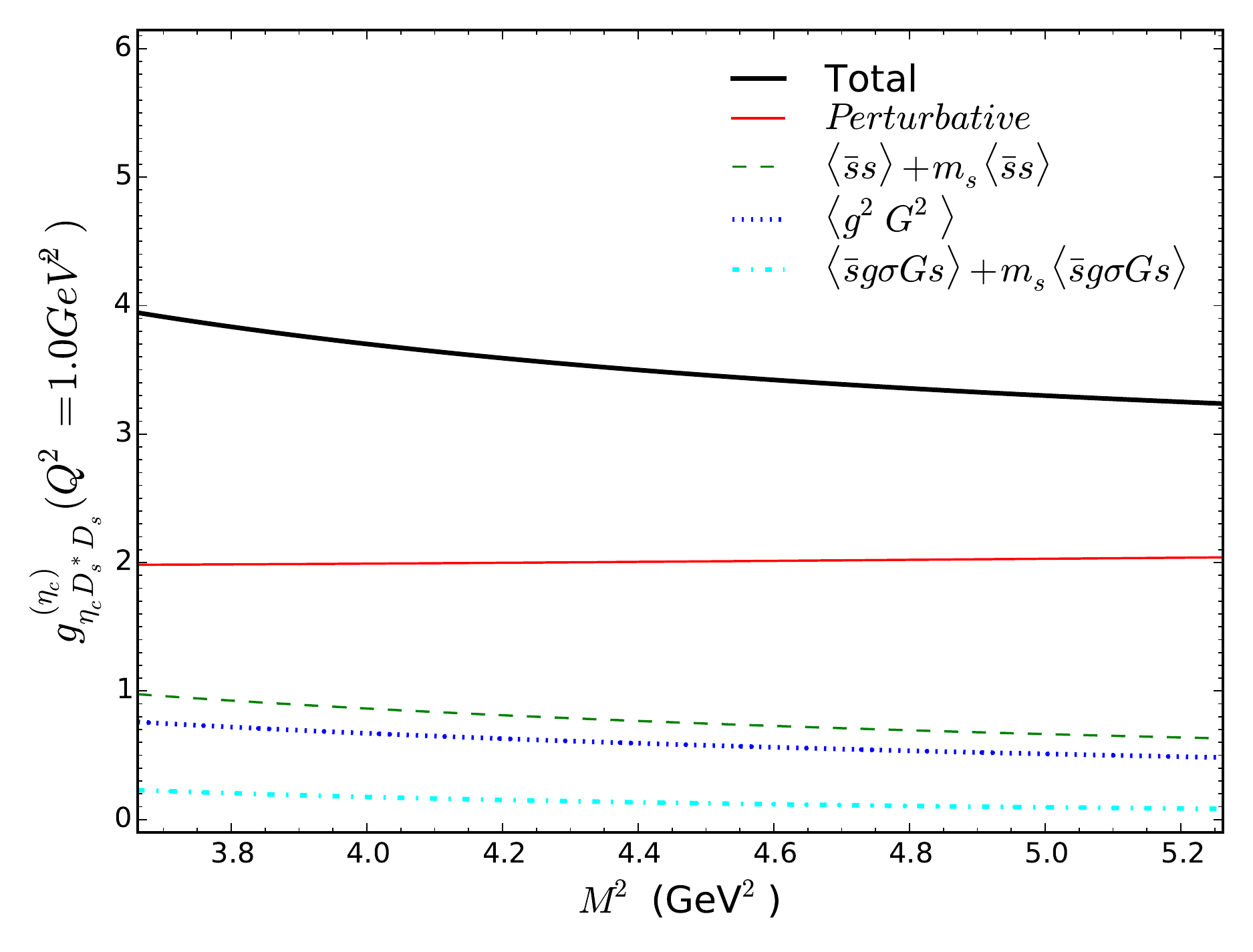}\label{subfigestabilidade:d}}
\subfigure[]{\includegraphics[width=0.49\linewidth]{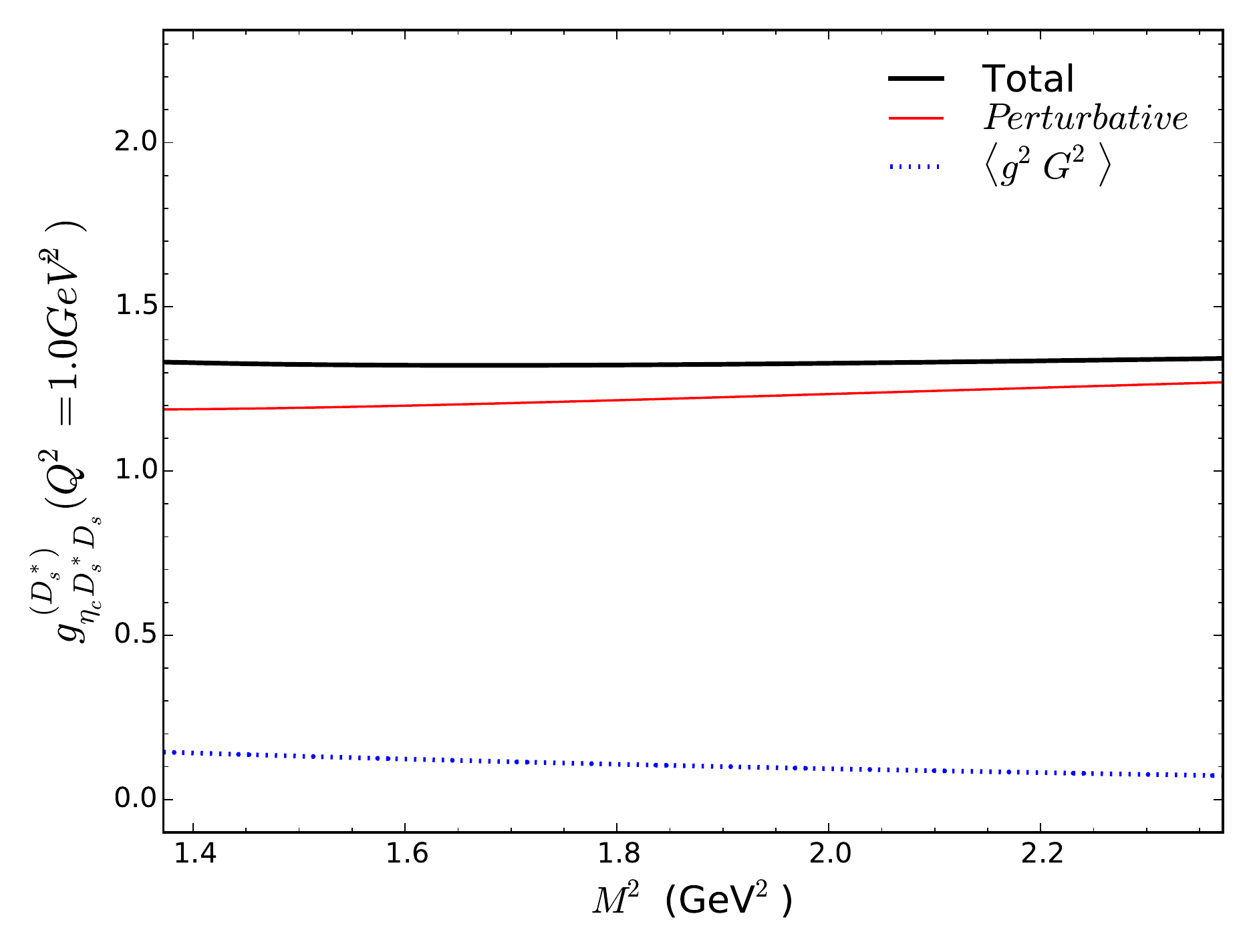}\label{subfigestabilidade:e}}
	%\subfigure[]{\includegraphics[width=0.49\linewidth]{figs/EtacDsEstDs/Contrib_Dspl.pdf}\label{subfigestabilidade:f}}
  \caption{\label{fig:estabilidade}  OPE  contributions  to  the  form
    factors  of the  vertices $\eta_c  D^* D$  for $\eta_c$  off-shell
    (panel \textbf{(a)}) and $D^*$  off-shell (panel \textbf{(b)}) and
    $\eta_c D_s^* D_s$ for $\eta_c$ off-shell (panel \textbf{(c)}) and
    $D_s^*$ off-shell (panel \textbf{(d)}).}
\end{figure}

The  continuum threshold  parameters  are  defined as  $s_0  = (m_i  +
\Delta_i)^2$  and $u_0  =  (m_o +  \Delta_o)^2$,  where the  quantities
$\Delta_i$  and  $\Delta_o$  are   determined  by  the  aforementioned
criterion of stability  of the QCDSR. The function  of such parameters
is to include the pole  contribution and to simultaneously exclude the
$h.r.$ contributions  of the QCDSR.   For this purpose, the  values of
$\Delta_{\eta_c}$, $\Delta_{\mathcal{D}^*}$ and $\Delta_{\mathcal{D}}$
should not be  very far from the experimental  values (when available)
of the difference between the pole  masses and the first excited state
of                   each                  meson~\cite{Agashe:2014kda,
  Badalian:2011tb,PhysRevD.87.052005}.        The      values       of
$\Delta_{\eta_c}$, $\Delta_{\mathcal{D}^*}$ and $\Delta_{\mathcal{D}}$
determined    in     our    analysis    were     $\Delta_{\eta_c}    =
\Delta_{\mathcal{D}}   =  0.6\,GeV$   and  $\Delta_{\mathcal{D}^*}   =
0.5\,GeV$. In the Fig.~\ref{fig:estabilidade},  it is possible  to see
that the use of these values leads to stable Borel windows for all the
off-shell  cases of  both  vertices.   The plot  for  the  case of  the
$\mathcal{D}$ off-shell  were omitted  from Fig.~\ref{fig:estabilidade}
due to the similarity with the $\mathcal{D}^*$ off-shell case.

In  Fig.~\ref{fig:estabilidade}, it is also possible to verify that
the perturbative term is in fact the leading term of the OPE, followed
by the quark condensates (only for  the case of the $\eta_c$ off-shell)
and the  gluon condensates. In the  same figure, it is  shown that the
contribution from the  terms $\langle \bar{s}g\sigma G  s \rangle$ are
small and could easily be neglected without significantly altering the
results.

With regard  of the choice  of the  tensorial structures used  in this
work, if  it was  possible to  use the complete  OPE series,  both the
structures $p_\mu$  and $p'_\mu$ from the  Eq.~(\ref{qhb}), would lead
to a valid QCDSR.  In the actual calculation however, in which the OPE
series  must  be truncated  at  some  order, some  approximations  are
necessary   to   deal  with   the   $h.r.$   terms  that   appear   in
phenomenological side.   It is  not possible,  therefore, to  use both
structures with equivalence.  We use  herein the $p_\mu$ structure for
the cases  in which  $\eta_c$ or  $\mathcal{D}^*$ are  off-shell while
using the $p'_\mu$ structure in the $\mathcal{D}$ off-shell case.  The
$p_\mu$ structure  for the  case of  $\mathcal{D}$ off-shell  does not
lead to coupling constants consistent  with the other off-shell cases.
The same applies  to the $p'_\mu$ structure of  the $\eta_c$ off-shell
case,  while for  the $\mathcal{D}^*$  off-shell case,  this structure
does not allow to obtain a valid Borel window.

\begin{figure}[!ht]
\centering
	\subfigure[]{\includegraphics[width=0.49\linewidth]{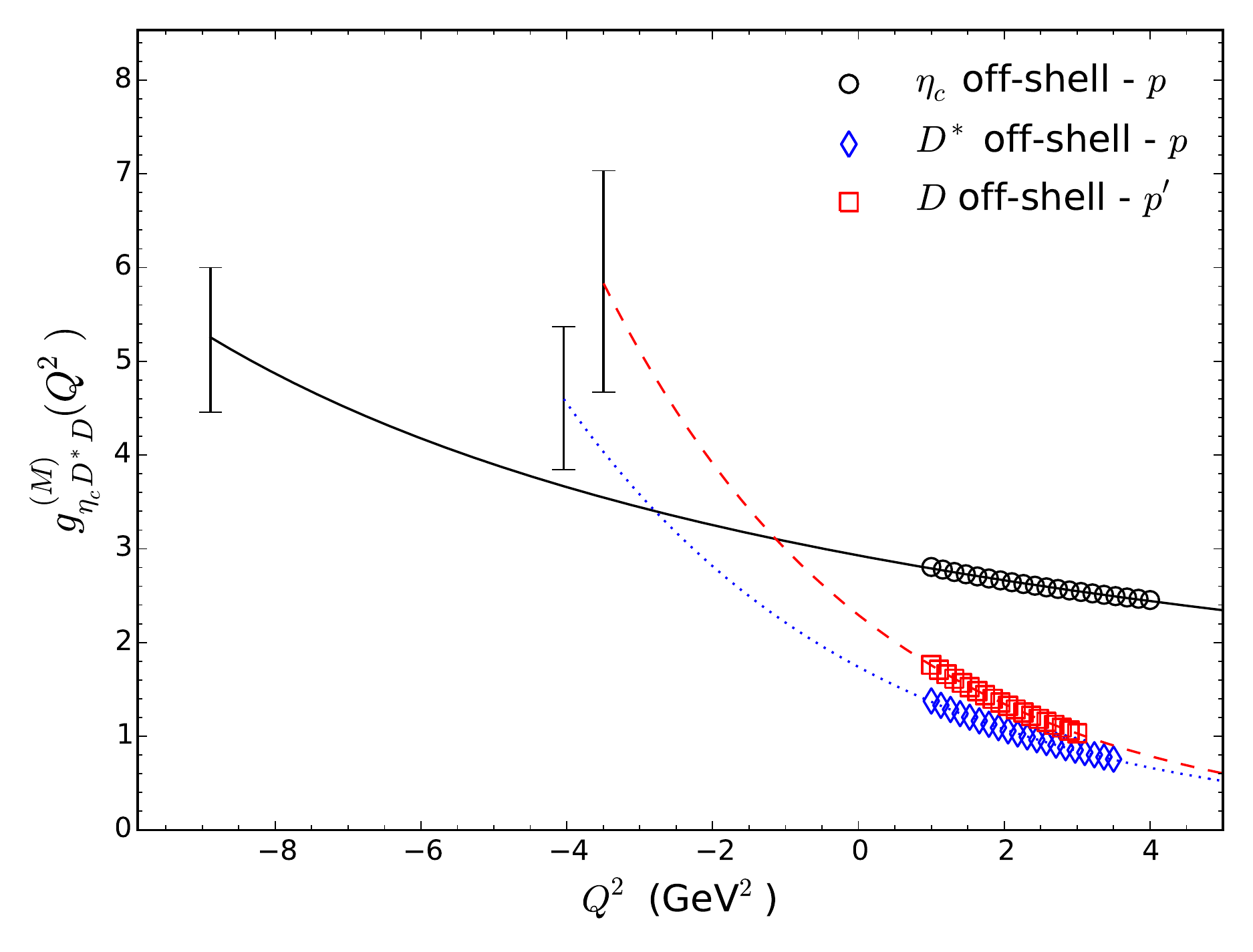}\label{subfigformfactor:a}}
	\subfigure[]{\includegraphics[width=0.49\linewidth]{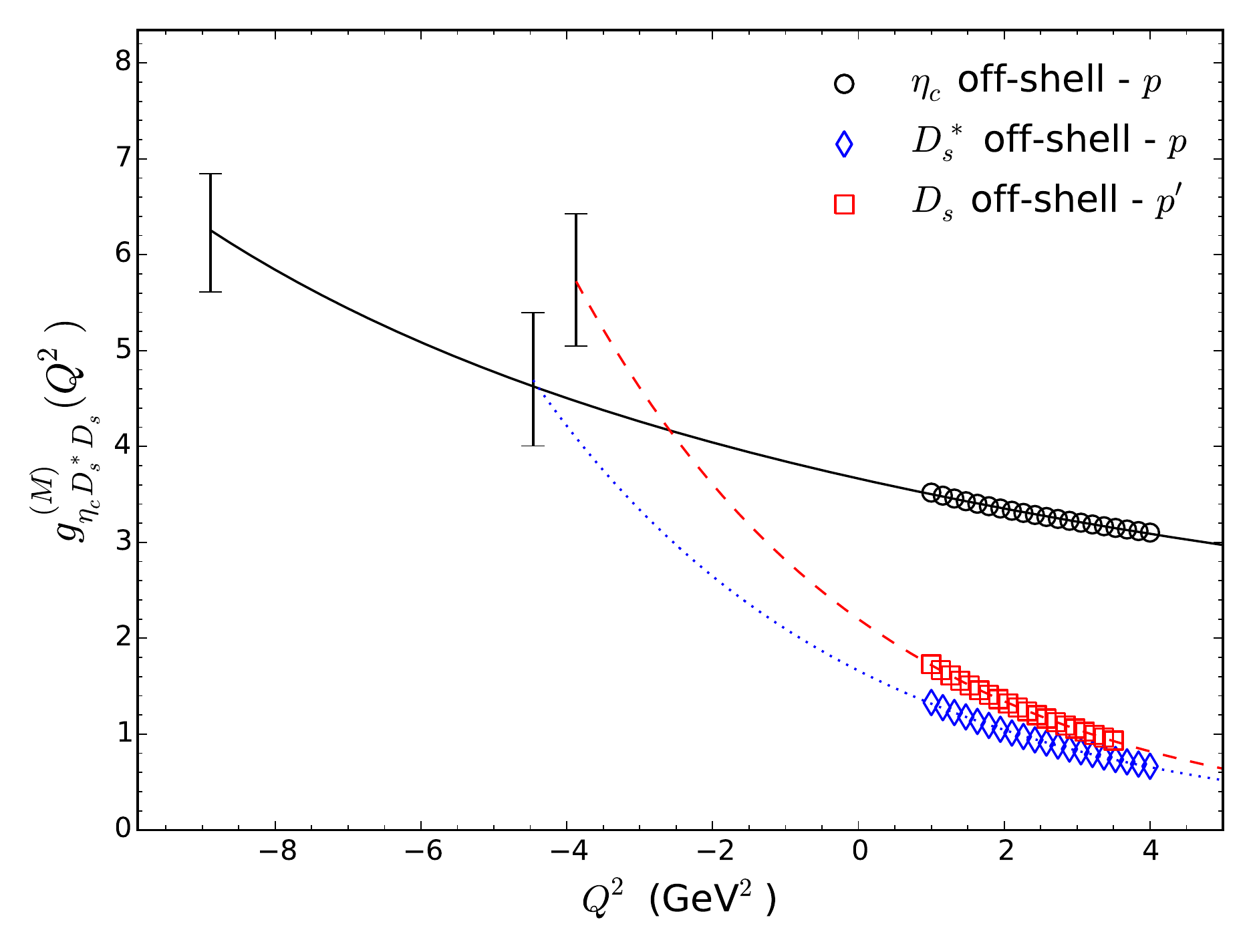}\label{subfigformfactor:b}}
  \caption{\label{fig:formfactors} Form factors of  the $\eta_c D^* D$
    vertex  (panel \textbf{(a)})  and  the $\eta_c  D_s^* D_s$  vertex
    (panel  \textbf{(b)}). Parametrizations  are  summarized in  Table
    \ref{tab:results}. The  coupling constants are represented  by the
    points with the error bars.}
\end{figure}

In  the  Table~\ref{tab:results}, it  is  presented  the form  factors
$g^{(M)}_{\eta_c            \mathcal{D}^*           \mathcal{D}}(Q^2)$
($M=\eta_c,\mathcal{D}^*,\mathcal{D}$)   obtained  herein   and  their
respective windows for $Q^2$ and $M^2$. It is also shown in this table
the coupling constants  $g_{\eta_c D^* D}$ and  $g_{\eta_c D_s^* D_s}$
obtained from these form factors with their error estimates.  In order
to obtain the form factors, the fit of the results were made using the
monopolar ($\frac{A}{B+Q^2}$) or exponential (${Ae^-Q^2/B}$) curves in
all cases,  for simplicity  and consistency  with our  previous works.
The form  factors for the  case $\eta_c$ off-shell were  well adjusted
for  both   monopolar  and  exponential  curves,   but  the  monopolar
adjustment presented  the coupling constants  more in line  with other
off-shell  cases.   The  cases of  $\mathcal{D}^*$  and  $\mathcal{D}$
off-shell  could only  be adjusted  by exponential  curves, while  the
monopolar adjustments  led to  divergences in  the calculation  of the
coupling constant.  The Fig.~\ref{fig:formfactors} shows the fits used
for  all three  off-shell  cases  of both  vertices  included in  this
study. The  coupling constants in  this figure are represented  by the
points with error  bars, where we see that for  each vertex, the three
off-shell cases lead to coupling  constants compatible with each other
within a confidence range of $ 1\sigma$.

\begin{table}[ht]
\caption{\label{tab:results}Parametrization  of the  form factors  and
  numerical  results  for the  coupling  constant  of this  work.  The
  calculation of $\sigma$ is explained in the text.}

\begin{tabular*}{\linewidth}{@{\extracolsep{\fill}} cccccc}
\hline
%& & \multicolumn{4}{c}{Quantity}\\
Vertex & Off-shell meson & $Q^2 (\text{GeV}^2)$ & $M^2 (\text{GeV}^2)$ & $g^{(M)}_{\eta_c \mathcal{D}^*\mathcal{D}}(Q^2)$ & $g^{(M)}_{\eta_c \mathcal{D}^*\mathcal{D}}\pm \sigma$ \\
\hline
\multirow{3}{*}{$\eta_c D^* D$} & $\eta_c$ & [1.0, 4.0] & [2.9, 4.0] & $\frac{58.79}{20.07+Q^2}$ & $5.25^{+0.75}_{-0.80}$  \\
& $D^*$ & [1.0, 3.5] & [1.2, 2.2] & $1.739\,e^{-Q^2/4.158}$  & $4.60^{+0.77}_{-0.75}$ \\
& $D$ & [1.0, 3.0] & [1.1, 2.1] & $2.294\, e^{-Q^2/3.747}$  & $5.83^{+1.20}_{-1.16}$ \\
\hline
\multirow{3}{*}{$\eta_c D_s^* D_s$} & $\eta_c$ & [1.0, 4.0] & [3.7, 5.3] & $\frac{78.67}{21.46+Q^2}$  & $6.25^{+0.59}_{-0.64}$ \\
& $D_s^*$ & [1.0, 4.0] & [1.4, 2.4] & $1.662\, e^{-Q^2/4.299}$  & $4.69^{+0.71}_{-0.69}$ \\
& $D_s$ & [1.0, 3.5] & [1.4, 2.4] & $2.200\, e^{-Q^2/4.051}$  &  $5.72^{+0.71}_{-0.68}$\\
\hline
\end{tabular*}
\end{table}

We   use    in   this    work   the    same   procedures    shown   in
Ref.~\cite{Rodrigues:2010ed,CerqueiraJr2012130} for  estimating errors
of coupling  constants.  This estimate  has been done by  studying the
coupling constant  behavior with the  individual variation of  each of
the  parameters   involved  in  the  calculations   within  their  own
uncertainties.   All parameters  considered in  the estimation  of the
coupling constants errors are shown in Table~\ref{tab:errors}, as well
as their values  and their uncertainties. The  masses, decay constants
and  condensates  have  their  own errors  from  experiments  or  from
theoretical calculations  in the  literature.  The uncertainty  due to
the Borel  mass $M^2$ was  computed by  the standard deviation  of the
form factor within the Borel window. The error due to the $Q^2$ window
was estimated by calculating the  effect that large variations in this
window (variations of  $\pm20\%$ in its width and its  upper and lower
limits)  has on  the constant  coupling. The  same was  done with  the
continuum threshold  parameters ($\Delta_i $, $\Delta_o  $), whose the
studied variation  was $\pm$0.1  GeV ($\sim20\%$) in  both parameters.
In the error estimations it was also taken into account variations in
the   fitting    parameters   of    the   form   factors    shown   in
Table~\ref{tab:results}.

Finally, we  calculate the  mean and standard  deviation of  all these
parameter variations.  In the  Table~\ref{tab:errors}, it is presented
the deviations percentage  of the coupling constant  due to individual
variation  of each  parameter for  the  two vertices  studied and  its
respective three off-shell cases.  In this  table we can see that most
of  the parameters  has little  impact on  the value  of the  coupling
constant   ($\Delta_{g_{\eta_c\mathcal{D}^*\mathcal{D}}}<10\%$).    We
believe that this good behavior makes unnecessary a more sophisticated
analysis of the parameters of QCDSR.

From the results presented  in Table~\ref{tab:results}, we compute
the mean value of  the coupling constants obtained for both
vertices and we obtain as our final results the following coupling constants:
\begin{align}
g_{\eta_c D^* D} = 5.23^{+1.80}_{-1.38}
\label{eq:cteetacdestd}\\
g_{\eta_c D_s^* D_s} = 5.55^{+1.29}_{-1.55}
\label{eq:cteetacdsestds}
\end{align}

\begin{table}[ht]
\caption{\label{tab:errors}Percentage   deviation   of  the   coupling
  constants  ($\Delta  g_{\eta_c \mathcal{D}^*  \mathcal{D}}$)  coming
  from the propagation of the error in each parameter.}
\begin{tabular*}{\linewidth}{@{\extracolsep{\fill}} ccccccc}\hline
 & \multicolumn{6}{c}{Deviation $\Delta g_{\eta_c \mathcal{D}^* \mathcal{D}}$ (\%)} \\
Vertex & \multicolumn{3}{c}{$\eta_c D^* D$} & \multicolumn{3}{c}{$\eta_c D_s^* D_s$} \\
 Parameter  / Off-shell meson & $\eta_c$ & $D^*$ & $D$ & $\eta_c$ & $D_s^*$ & D\\ 
\hline
$f_{\eta_c} = 394.7 \pm 2.4$ (MeV)~\cite{PhysRevD.82.114504}     &  0.50  &  0.50 & 0.50 & 0.50 & 0.50 & 0.50 \\
$f_{D^*} = 242^{+20}_{-12}$ (MeV)~\cite{PhysRevD.88.014015}     & 5.32 & 5.32 & 5.33 & -- & -- & --   \\
$f_{D} = 206.7 \pm 8.5\pm 2.5$   (MeV)~\cite{PhysRevD.86.010001}   & 4.35 & 4.36 & 4.35  & -- & -- & -- \\
$f_{D_s^*} = 301 \pm 13$ (MeV)~\cite{PhysRevD.60.074501,PhysRevD.75.116001} & -- & -- & -- & 3.53 & 3.53 & 3.53  \\
$f_{D_s} = 257.5 \pm 6.1$    (MeV)~\cite{Nakamura:2010zzi}  & -- & -- & --  & 1.93 & 1.93 & 1.93 \\
$m_c = 1.27^{+0.07}_{-0.09}$ (GeV)~\cite{Nakamura:2010zzi} & 3.90 & 3.33 & 6.64 & 3.38 & 2.74 & 5.89 \\
$m_s = 101^{+29}_{-21}$ (MeV)~\cite{Nakamura:2010zzi}      & -- & -- & --  & 3.61 & 0.40 & 3.24 \\ 
$M^2$ (MeV$^2$)$^{(a)}$    & 7.57 & 13.55 & 16.97 & 1.57 & 12.55 & 7.82  \\
$\Delta_i \pm 0.1$ (GeV),$\Delta_o \pm 0.1$ (GeV)  & 6.43 & 4.63 & 3.14 & 5.25 & 4.82 & 3.06 \\
$Q^2 \pm 20\%$ (GeV$^2$)$^{(a)}$  & 4.26 & 2.11 & 1.03 & 3.61 & 3.27 & 2.06 \\
$\langle \bar{u}u\rangle = \langle \bar{d}d\rangle  = -(230 \pm 30)^3$ (MeV$^3$)~\cite{Matheus:2005yu,Bracco:2004rx} & 2.35 &  --  &  -- & --  &  -- &  --  \\
$\langle \bar{s}s\rangle = -(290 \pm 15)^3$ (MeV$^3$)~\cite{PhysRevD.87.034503} & -- & -- & -- & 1.24  & -- & -- \\
$\langle g^2 G^2 \rangle = 0.88\pm 0.16$(GeV$^4$)~\cite{Narison2012412} & 2.21 & 0.61 & 2.58 & 0.51 & 0.47 & 2.93 \\
$\langle \bar{q}g\sigma \cdot G q \rangle = (0.8 \pm 0.2) \langle \bar{q}q\rangle $ (GeV$^5$)~\cite{Ioffe2006232} & 3.63 &  --  &  -- &  1.33   &  -- &  -- \\
%$\langle \bar{s}g\sigma \cdot G s \rangle = (0.8\pm 0.2)\langle \bar{s}s\rangle$(GeV$^5$)~\cite{Ioffe2006232} & -- & -- & -- &  1.33  &  -- &  --    \\
Fitting parameters  & 3.84 & 0.19 & 0.09 & 3.31 & 0.28 & 0.16\\ \hline
\end{tabular*}
\footnotetext[1]{The intervals for these quantities are those of Table~\ref{tab:results}.}
\end{table}

\section{Conclusions}

In this work, we have  obtained the  constant coupling of  the charmed
meson vertices  $\eta_c D^*  D$ and $\eta_c  D_s^* D_s$,  applying the
QCDSR formalism for three different off-shell mesons. The advantage of
this  method  is   the  minimization  of  the   uncertainties  of  the
calculations.

The   numerical   results   obtained for the coupling constants  are:
\begin{align*}
g_{\eta_c D^* D} = 5.23^{+1.80}_{-1.38},\\
g_{\eta_c  D_s^* D_s}  = 5.55^{+1.29}_{-1.55}.
\end{align*}
The  difference  between  the  values of  the  coupling  constants  is
expected due  to the   breaking of  the $SU(4)$  symmetry. The
effect is  an increase  of about  $6\%$ in the  value of  the coupling
constant  when the  mass of  the  strange-quark is  introduced in  the
calculations of the correlation functions.
The parametrization  of the form  factors are similar to  our previous
works.  The  monopolar parametrization  works for  the cases  with the
heaviest  meson off-shell  ($\eta_c$), while  the gaussian  exponential
works for the lightest ones are off-shell.

We  can also  compare  our  results for  the  coupling constants  with
previous calculations, that are shown in Table~\ref{tab:compare}.

\begin{table}[ht]
\caption{\label{tab:compare}Values of the coupling constants computed with different approaches: Vector Meson Dominance (VMD) and relativistic Constituent Quark Model (CQM).}

\begin{tabular*}{0.8\linewidth}{@{\extracolsep{\fill}} cccc}
  \hline
Method \& References &  $g_{\eta_c D^* D}$ & $g_{\eta_c  D_s^* D_s}$ \\
\hline
This work                      &  $5.23^{+1.80}_{-1.38}$ &  $5.55^{+1.29}_{-1.55}$\\
VMD~\cite{Wang:2012wj}         & $7.68$               &    --           \\ 
VMD~\cite{Zhang:2008ab}        & $7.44$               &    --            \\ 
relativistic CQM~\cite{PhysRevD.93.016004} &  $15.51\pm0,45$      &  $14.15\pm0.52$ \\
QCDSR and SU(4)~\cite{Matheus:2005yu,OsorioRodrigues2014143} & $5.8 \pm 0.8$ & $5.98^{+0.67}_{-0.58}$\\
\hline
\end{tabular*}
\end{table}

The comparison  with the coupling  constants obtained from  chiral and
heavy  quark  limit relation  (HQL)  combined  with the  vector  meson
dominance  (VMD) in  Refs.~\cite{Wang:2012wj,Zhang:2008ab} shows  that
the values  are compatible within  the errors. The  difference between
them for our  value are of approximately $30\%$.   However, the values
obtained in Ref.~\cite{PhysRevD.93.016004}  using relativistic QCM are
much bigger than our results and those from VMD, about $250-300\%$ for
the mean values. This approach the coupling constant is obtained using
a Gaussian  parametrization.  In  our case, QCDSR  method, we  use the
convergence  of three  different form  factors of  the same  vertex to
obtain the coupling constant.  The  QCDSR method thus reduce the erros
derived from the parametrization.

If  we compare  our results  of the  coupling constants  using $SU(4)$
relations,  $g_{\eta_c D^*  D}$ and  $g_{\eta_c D_s^*  D_s}$ with  our
previous  QCDSR  results  for  the  couplings  $g_{J/\psi  D  D}$  and
$g_{J/\psi  D_s  D_s}$  ~\cite{Bracco:2011pg}.  There  are  compatibility
within $1\sigma$ and varying approximately $10\%$ from each other.

\section*{Acknowledgments}
This work has been supported by CNPq and FAPERJ. 

\bibliography{bibliografia}

\appendix
\label{appendix:remaningexpressions}
\section{}

Here we present the full expressions for the contributions coming from the condensates $\langle g^2 G^2\rangle$ and $\langle \bar{q}g\sigma G q \rangle$ to the correlator of Eq.~(\ref{eq:nonpertcontrib}) in the $\eta_c$ off-shell case.

\begin{eqnarray}
\mathcal{B} \mathcal{B} \left[\Gamma^{\langle g^2G^2 \rangle}_{{p}_\mu} \right ] &=& 
 \frac{\langle g^2G^2 \rangle}{96\pi^2} \int^\infty_{1/M^2} d\alpha e^{\frac{t(\alpha M'^2-1)}{M^2+M'^2} + \frac{\alpha \left (1 +\frac{M'^2}{M^2} \right )\left [ m_c^2- m_s^2 - \alpha^2 m_c^2M'^2\right]}{\alpha M'^2-1}  }  \left (  F_{(d)} +  F_{(e)} \right. \nonumber \\  && \left.+  F_{(f)} + F_{(g)} + F_{(h)} + F_{(i)} \right ){p}_\mu
\end{eqnarray}

\begin{eqnarray}
F_{(d)} &=& (M'^6 (\alpha^2 M'^2 m_c m_q^2 M^4+\alpha^3 M'^4 m_c^2 m_q M^4-\alpha^2 M'^2 m_c^2 m_q M^4-3 \alpha^2 M'^4 m_q M^4+6 \alpha M'^2 m_q M^4 \nonumber\\
&&-3 m_q M^4+4 \alpha^2 M'^4 m_c M^4-7 \alpha M'^2 m_c M^4+3 m_c M^4+\alpha^2 M'^4 m_c m_q^2 M^2+\alpha M'^2 m_c m_q^2 M^2 \nonumber\\
&&+\alpha^3 M'^6 m_c^2 m_q M^2-\alpha^2 M'^4 m_c^2 m_q M^2-\alpha^3 M'^6 m_c^3 M^2+2 \alpha^2 M'^4 m_c^3 M^2-\alpha M'^2 m_c^3 M^2 \nonumber\\
&& +2 \alpha^2 M'^6 m_c M^2-3 \alpha M'^4 m_c M^2+M'^2 m_c M^2+\alpha M'^4 m_c m_q^2-\alpha^3 M'^8 m_c^3+2 \alpha^2 M'^6 m_c^3\nonumber\\
&&-\alpha M'^4 m_c^3))/((\alpha M'^2-1) M^6 (M^2+M'^2)^2)
\end{eqnarray}

\begin{eqnarray}
F_{(e)} &=& (M'^2 (2 \alpha^2 m_q^3 M^8+2 \alpha^3 M'^2 m_c m_q^2 M^8-2 \alpha^2 m_c m_q^2 M^8+2 \alpha^4 M'^6 m_q M^8-8 \alpha^3 M'^4 m_q M^8\nonumber\\
&&+{10} \alpha^2 M'^2 m_q M^8-4 \alpha m_q M^8-\alpha^4 M'^6 m_c M^8+5 \alpha^3 M'^4 m_c M^8-7 \alpha^2 M'^2 m_c M^8+3 \alpha m_c M^8\nonumber\\
&&+2 \alpha^5 M'^8 m_q t M^6-6 \alpha^4 M'^6 m_q t M^6+8 \alpha^3 M'^4 m_q t M^6-6 \alpha^2 M'^2 m_q t M^6+2 \alpha m_q t M^6\nonumber\\
&&+4 \alpha^4 M'^6 m_c t M^6-6 \alpha^3 M'^4 m_c t M^6+4 \alpha^2 M'^2 m_c t M^6-\alpha m_c t M^6-2 \alpha^4 M'^6 m_q^3 M^6\nonumber\\
&&+4 \alpha^3 M'^4 m_q^3 M^6+4\alpha^2 M'^2 m_q^3 M^6+2 \alpha m_q^3 M^6+\alpha^4 M'^6 m_c m_q^2 M^6+4 \alpha^3 M'^4 m_c m_q^2 M^6\nonumber\\
&&-5 \alpha^2 M'^2 m_c m_q^2 M^6-2 \alpha^5 M'^8 m_c^2 m_q M^6+6 \alpha^4 M'^6 m_c^2 m_q M^6-8 \alpha^3 M'^4 m_c^2 m_q M^6\nonumber\\
&&+6 \alpha^2 M'^2 m_c^2 m_q M^6-2 \alpha m_c^2 m_q M^6-4 \alpha^3M'^6 m_q M^6+4 \alpha^2 M'^4 m_q M^6+4 \alpha M'^2 m_q M^6\nonumber\\
&&-4 m_q M^6+\alpha^5 M'^8 m_c^3 M^6-3 \alpha^4 M'^6 m_c^3 M^6+3 \alpha^3 M'^4 m_c^3 M^6-\alpha^2 M'^2 m_c^3 M^6+4 \alpha^3 M'^6 m_c M^6\nonumber\\
&&-8 \alpha^2 M'^4 m_c M^6+4 \alpha M'^2 m_c M^6+2 \alpha^4 M'^8 m_q t M^4-2 \alpha^3 M'^6 m_q t M^4-2 \alpha^2 M'^4 m_q t M^4\nonumber\\
&&+2 \alpha M'^2 m_q t M^4-2 \alpha^4 M'^8 m_c t M^4+6 \alpha^3 M'^6 m_c t M^4-6 \alpha^2 M'^4 m_c t M^4+2 \alpha M'^2 m_c t M^4\nonumber\\
&&-4 \alpha^4 M'^8 m_q^3 M^4+8 \alpha^3 M'^6 m_q^3 M^4+2 \alpha^2 M'^4 m_q^3 M^4+6 \alpha M'^2 m_q^3 M^4+2 \alpha^4 M'^8 m_c m_q^2 M^4\nonumber\\
&&+2 \alpha^3 M'^6 m_c m_q^2 M^4-4 \alpha^2 M'^4 m_c m_q^2 M^4-4 \alpha^5 M'^{10} m_c^2 m_q M^4+12 \alpha^4 M'^8 m_c^2 m_q M^4\nonumber\\
&& -18 \alpha^3 M'^6 m_c^2 m_q M^4+16 \alpha^2 M'^4 m_c^2 m_q M^4-6 \alpha M'^2 m_c^2 m_q M^4-2 \alpha^4 M'^{10} m_q M^4\nonumber\\
&& +4 \alpha^3 M'^8 m_q M^4-{10} \alpha^2 M'^6 m_q M^4+16 \alpha M'^4 m_q M^4-8 M'^2 m_q M^4+2 \alpha^5 M'^{10} m_c^3 M^4\nonumber\\
&& -6 \alpha^4 M'^8 m_c^3 M^4+6 \alpha^3 M'^6 m_c^3 M^4-2 \alpha^2 M'^4 m_c^3 M^4+\alpha^4 M'^{10} m_c M^4-\alpha^3 M'^8 m_c M^4\nonumber\\
&& -\alpha^2 M'^6 m_c M^4+\alpha M'^4 m_c M^4+2 \alpha^3 M'^8 m_q t M^2-4 \alpha^2 M'^6 m_q t
 M^2+2 \alpha M'^4 m_q t M^2\nonumber\\
&& -2 \alpha^4 M'^{10} m_c t M^2+6 \alpha^3 M'^8 m_c t M^2-6 \alpha^2 M'^6 m_c t M^2+2 \alpha M'^4 m_c t M^2-2 \alpha^4 M'^{10} m_q^3 M^2\nonumber\\
&& +4 \alpha^3 M'^8 m_q^3 M^2+6 \alpha M'^4 m_q^3 M^2+\alpha^4 M'^{10} m_c m_q^2 M^2-\alpha^2 M'^6 m_c m_q^2 M^2\nonumber\\
&& -2 \alpha^5 M'^12 m_c^2 m_q M^2+6 \alpha^4 M'^{10} m_c^2 m_q M^2-12 \alpha^3 M'^8 m_c^2 m_q M^2+14 \alpha^2 M'^6 m_c^2 m_q M^2\nonumber\\
&& -6 \alpha M'^4 m_c^2 m_q M^2-4 \alpha^2 M'^8 m_q M^2+8 \alpha M'^6 m_q M^2-4 M'^4 m_q M^2+\alpha^5 M'^12 m_c^3 M^2\nonumber\\
&& -3 \alpha^4 M'^{10} m_c^3 M^2+3 \alpha^3 M'^8 m_c^3 M^2- \alpha^2 M'^6 m_c^3 M^2+2 \alpha M'^6 m_q^3-2 \alpha^3 M'^{10} m_c^2 m_q\nonumber\\
&& +4 \alpha^2 M'^8 m_c^2 m_q-2 \alpha M'^6 m_c^2 m_q-\alpha^5 M'^8 m_c t M^6))/(2 (\alpha M'^2-1)^2 (M^2+M'^2)^4)
\end{eqnarray}

\begin{eqnarray}
F_{(f)} &=& (M'^2 (M^2+M'^2) (\alpha^2 m_q^3 M^4+\alpha^3 M'^2 m_c m_q^2 M^4-\alpha^2 m_c m_q^2 M^4-2 \alpha^2 M'^2 m_q M^4+2 \alpha m_q M^4\nonumber\\
&&-3 \alpha^3 M'^4 m_c M^4+6 \alpha^2 M'^2 m_c M^4-3 \alpha m_c M^4+\alpha^2 M'^2 m_q^3 M^2+\alpha m_q^3 M^2 +\alpha^3 M'^4 m_c m_q^2 M^2 \nonumber\\
&&-\alpha^2 M'^2 m_c m_q^2 M^2-\alpha^3 M'^4 m_c^2 m_q M^2+2 \alpha^2 M'^2 m_c^2 m_q M^2-\alpha m_c^2 m_q M^2 -\alpha^2 M'^4 m_q M^2 +m_q M^2 \nonumber\\
&&+\alpha M'^2 m_q^3-\alpha^3 M'^6 m_c^2 m_q+2 \alpha^2 M'^4 m_c^2 m_q-\alpha M'^2 m_c^2 m_q))/((\alpha M'^2-1)^4 M^6)
\end{eqnarray}

\begin{eqnarray}
F_{(g)} &=&  (M'^2 (3 \alpha^2 m_q^2 M^8+6 \alpha^3 M'^2 m_c m_q M^8-6 \alpha^2 m_c m_q M^8+\alpha^3 M'^4 M^8-2 \alpha^2 M'^2 M^8+\alpha M^8-M^6\nonumber\\
&&+\alpha^4 M'^6 t M^6-2 \alpha^3 M'^4 t M^6+\alpha^2 M'^2 t M^6+8 \alpha^2 M'^2 m_q^2 M^6+3 \alpha m_q^2 M^6+15 \alpha^3 M'^4 m_c m_q M^6\nonumber\\
&&-12 \alpha^2 M'^2 m_c m_q M^6-3 \alpha m_c m_q M^6-\alpha^4 M'^6 m_c^2 M^6-\alpha^3 M'^4 m_c^2 M^6+5 \alpha^2 M'^2 m_c^2 M^6-2 M'^4 M^2\nonumber\\
&&-3 \alpha m_c^2 M^6-\alpha^2 M'^4 M^6+2 \alpha M'^2 M^6+\alpha^3 M'^6 t M^4-2 \alpha^2 M'^4 t M^4+\alpha M'^2 t M^4+7 \alpha^2 M'^4 m_q^2 M^4\nonumber\\
&&+8 \alpha M'^2 m_q^2 M^4+12 \alpha^3 M'^6 m_c m_q M^4-6 \alpha^2 M'^4 m_c m_q M^4-6 \alpha M'^2 m_c m_q M^4-2 \alpha^4 M'^8 m_c^2 M^4\nonumber\\
&&-4 \alpha^3 M'^6 m_c^2 M^4+14 \alpha^2 M'^4 m_c^2 M^4-8 \alpha M'^2 m_c^2 M^4-\alpha^3 M'^8 M^4-\alpha^2 M'^6 M^4+5 \alpha M'^4 M^4\nonumber\\
&&+2 \alpha^2 M'^6 m_q^2 M^2+7 \alpha M'^4 m_q^2 M^2+3 \alpha^3 M'^8 m_c m_q M^2-3 \alpha M'^4 m_c m_q M^2-\alpha^4 M'^{10} m_c^2 M^2\nonumber\\
&&-5 \alpha^3 M'^8 m_c^2 M^2+13 \alpha^2 M'^6 m_c^2 M^2-7 \alpha M'^4 m_c^2 M^2-2 \alpha^2 M'^8 M^2+4 \alpha M'^6 M^2+2 \alpha M'^6 m_q^2\nonumber\\
&&-3 M'^2 M^4 -2 \alpha^3 M'^{10} m_c^2+4 \alpha^2 M'^8 m_c^2-2 \alpha M'^6 m_c^2))
/(\alpha (\alpha M'^2-1)^2 M^2 (M^2+M'^2)^3)
\end{eqnarray}

\begin{eqnarray}
F_{(h)} &=& (M'^2 (\alpha^2 M'^2 m_q^2 M^6-3 \alpha^2 M'^2 m_c m_q M^6+\alpha M'^2 M^6-M^6-\alpha^3 M'^6 t M^4+2 \alpha^2 M'^4 t M^4\nonumber\\
&&-\alpha M'^2 t M^4+2 \alpha^2 M'^4 m_q^2 M^4+2 \alpha M'^2 m_q^2 M^4+6 \alpha^3 M'^6 m_c m_q M^4-6 \alpha^2 M'^4 m_c m_q M^4\nonumber\\
&&-5 \alpha^3 M'^6 m_c^2 M^4+10 \alpha^2 M'^4 m_c^2 M^4-5 \alpha M'^2 m_c^2 M^4+3 \alpha M'^4 M^4-3 M'^2 M^4+\alpha^2 M'^6 m_q^2 M^2\nonumber\\
&&+4 \alpha M'^4 m_q^2 M^2+3 \alpha^3 M'^8 m_c m_q M^2-3 \alpha^2 M'^6 m_c m_q M^2-7 \alpha^3 M'^8 m_c^2 M^2+14 \alpha^2 M'^6 m_c^2 M^2\nonumber\\
&&-7 \alpha M'^4 m_c^2 M^2+2 \alpha M'^6 M^2-2 M'^4 M^2+2 \alpha M'^6 m_q^2-2 \alpha^3 M'^{10} m_c^2+4 \alpha^2 M'^8 m_c^2\nonumber\\
&&-2 \alpha M'^6 m_c^2+3 \alpha^3 M'^4 m_c m_q M^6))/(\alpha (\alpha M'^2-1) M^2 (M^2+M'^2)^3)
\end{eqnarray}

\begin{eqnarray}
F_{(i)} &=& (M'^2 (\alpha^2 M'^2 m_q^2 M^8+3 \alpha^3 M'^4 m_c m_q M^8-3 \alpha^2 M'^2 m_c m_q M^8+2 \alpha^3 M'^6 M^8-5 \alpha^2 M'^4 M^8-M'^2 M^6\nonumber\\
&&+4 \alpha M'^2 M^8-M^8+\alpha^4 M'^8 t M^6-3 \alpha^3 M'^6 t M^6+3 \alpha^2 M'^4 t M^6-\alpha M'^2 t M^6+2 \alpha^2 M'^4 m_q^2 M^6\nonumber\\
&&+2 \alpha M'^2 m_q^2 M^6+6 \alpha^3 M'^6 m_c m_q M^6-3 \alpha^2 M'^4 m_c m_q M^6-3 \alpha M'^2 m_c m_q M^6-\alpha^4 M'^8 m_c^2 M^6\nonumber\\
&&+3 \alpha^2 M'^4 m_c^2 M^6-2 \alpha M'^2 m_c^2 M^6+2 \alpha^3 M'^8 M^6-5 \alpha^2 M'^6 M^6+4 \alpha M'^4 M^6-4 \alpha M'^6 m_c^2 M^2\nonumber\\
&&+7 \alpha^2 M'^8 m_c^2 M^2+\alpha^2 M'^6 m_q^2 M^4+5 \alpha M'^4 m_q^2 M^4+3 \alpha^3 M'^8 m_c m_q M^4+3 \alpha^2 M'^6 m_c m_q M^4\nonumber\\
&&-6 \alpha M'^4 m_c m_q M^4-2 \alpha^4 M'^{10} m_c^2 M^4-\alpha^3 M'^8 m_c^2 M^4+8 \alpha^2 M'^6 m_c^2 M^4-5 \alpha M'^4 m_c^2 M^4\nonumber\\
&&+4 \alpha M'^6 m_q^2 M^2+3 \alpha^2 M'^8 m_c m_q M^2-3 \alpha M'^6 m_c m_q M^2-\alpha^4 M'^{12} m_c^2 M^2-2 \alpha^3 M'^{10} m_c^2 M^2\nonumber\\
&&+\alpha M'^8 m_q^2-\alpha^3 M'^{12} m_c^2+2 \alpha^2 M'^{10} m_c^2-\alpha M'^8 m_c^2))/(\alpha (\alpha M'^2-1)^2 M^4 (M^2+M'^2)^3)
\end{eqnarray}

\begin{eqnarray}
\mathcal{B} \mathcal{B} \left[\Gamma^{\langle \bar{q}g\sigma G q \rangle}_{{p}_\mu} \right ] 
&=& \frac{m_c\langle \bar{q}g\sigma G q \rangle }{4M^2M'^2}{p}_\mu e^{-\frac{m_c^2}{M^2}}e^{-\frac{m_c^2}{M'^2}} \left [ (m_c^2+M'^2)M^4 + (m_c^2-t)M^2M'^2 \right.\nonumber\\
&&\left. + m_c^2 M'^4 \right ]
\end{eqnarray}

\begin{eqnarray}
\mathcal{B} \mathcal{B} \left[\Gamma^{m_q\langle \bar{q}g\sigma G q \rangle}_{{p}_\mu} \right ] &=& -\frac{m_q\langle \bar{q}g\sigma G q \rangle}{24M'^6M^6}{p}_\mu e^{-2\frac{m_c^2}{M^2}} \left (2 m_c^4 M^6+33 M'^2 m_c^2 M^6-40 M'^4 M^6-6 M'^2 m_c^2 t M^4\right. \nonumber \\
&&-10 M'^4 t M^4+6 M'^2 m_c^4 M^4+10 M'^4 m_c^2 M^4-3 M'^6 t M^2+6 M'^4 m_c^4 M^2\nonumber \\
&&\left.+15 M'^6 m_c^2 M^2-2 M'^6 m_c^2 t+2 M'^6 m_c^4 -6 M'^6 M^4 \right )
\end{eqnarray}

\end{document}